\documentclass[
    prl,            
    reprint,        
    superscriptaddress, 
    showpacs,       
    floatfix,       
    amsmath,        
    amssymb         
]{revtex4-2}

\usepackage{graphicx}     

\usepackage[table]{xcolor}
\usepackage{dcolumn}      
\usepackage{bm}           
\usepackage{xcolor}       
\usepackage{easy-todo}
\usepackage{tikz}

\begin{document}





\title{Disentangling Entropic, Active, and Frictional Forces in Cytoskeletal Crosslinking}

\author{C\'edrik Barutel}
\affiliation{Institute of Applied Physics, TU Wien, Lehargasse 6, Vienna, 1060, Austria}
\author{Sebastian F\"urthauer}
\email{fuerthauer@iap.tuwien.ac.at}
\affiliation{Institute of Applied Physics, TU Wien, Lehargasse 6, Vienna, 1060, Austria}


\begin{abstract}
The forces that mixtures of motorized and passive crosslinking proteins collectively generate between cytoskeletal filaments within our cells are the key drivers of active cellular mechanics. Despite their importance, a unified theory to describe such crosslinking forces has so far been missing. 
In this paper, we derive a theory that predicts the forces generated collectively by crosslinking proteins linking two biopolymer filaments from measurable filament and crosslinker properties, using out-of-equilibrium thermodynamics.
Our framework allows us to decompose the forces generated by crosslinkers into three separate components: entropic, active, and frictional. In doing so, it offers a clear physical interpretation of the fundamental mechanisms by which crosslinking proteins self-organize and collectively generate forces.
We demonstrate the robustness and utility of this framework by applying it to four different experiments that probe the combined roles of passive and motorized crosslinkers.
For each experiment, our theoretical approach allows us to disentangle the relative contributions of entropic, active, and frictional forces, clarifying how different physical processes underpin collective force production.
In turn, this makes it possible to quantitatively compare and predict how various crosslinker combinations influence force generation between filaments, pattern formation along filaments, and the dynamics of filament pairs.
\end{abstract}

\maketitle

\section{Significance}
The ability of cells to mechanically deform, migrate, and divide relies on networks of biopolymer filaments that are interconnected by specialized proteins, i.e. the cytoskeleton. Crosslinking proteins can be either passive or motor-driven. Forces between cytoskeletal filaments emerge from the combined action of many such crosslinkers. Despite their crucial role in cell mechanics, a general theoretical framework describing the collective behavior and force generation by crosslink mixtures has been lacking. We present a unified theory that quantitatively predicts the force generated by mixtures of crosslinkers between cytoskeletal filaments. We show that these forces can be decomposed into three fundamental contributions (entropic, active, and frictional) and offer a physically intuitive framework that predicts their values starting from crosslink and filament properties.

\section{Introduction}
Cells move, divide, and rearrange. Their motion is powered by networks of biopolymer filaments (i.e. actin and microtubules) that are transiently connected by passive and motorized crosslinkers~\cite{alberts2002molecular, bray2000cell, howard2002mechanics}. A quantitative understanding of this material - the cytoskeleton - from fundamental physical principles is a central goal in biology and in the physics of living matter. While much progress has been made in predicting the material properties of cytoskeletal networks from given crosslinker-mediated filament interaction forces~\cite{liverpool2005bridging,saintillan2008instabilities,furthauer2019self, furthauer2021design,furthauer2022cross, de2024supramolecular}, how these interactions are controlled by the collective behavior of passive crosslinkers and motor molecules remains to be understood. We address this knowledge gap.

Crosslinker-mediated forces between cytoskeletal filaments have been experimentally measured using magnetic or optical traps for a wide variety of crosslinkers~\cite{Vale1985identification,Svoboda1994force,Carter2005mechanics,Hunt1994force,Finer1994single,Bormuth2009friction,Forth2014Asymetric,kuvcera2021anillin,Geyer2023horizontal,Alfieri2021two,Lansky2015adiffusible,Shimamoto2015measuring}. At the single-molecule level, the mechanical and kinetic properties of many crosslinkers and motors have been characterized~\cite{Vale1985identification,Svoboda1994force,Carter2005mechanics,Hunt1994force,Finer1994single,Bormuth2009friction}. Similar experiments have also probed the collective behavior of specific motor and crosslinker systems~\cite{Forth2014Asymetric,kuvcera2021anillin,Geyer2023horizontal,Alfieri2021two,Lansky2015adiffusible,Shimamoto2015measuring}. In general, the resulting force–velocity relationships depend on crosslinker concentrations, relative filament orientation, and the molecular composition of the surrounding medium, and are thus highly non-trivial. Here, we present a theory to explain these complexities.

The complex phenomenology seen in experiment has inspired numerical simulations using agent based models~\cite{Parmeggiani2004totally,Johann2012,Leduc2012,Descovich2018crosslinkers,LeraRamirez2019,Wierenga2020,Fiorenza2021,Striebel2022Regulation,Wen2022toward}. 
These simulations provide intriguing insight into biologically important questions. Examples are the generation of stable overlap regions in spindle midzones~\cite{LeraRamirez2019}, and collective behaviors of crosslink mixtures between filaments~\cite{Striebel2022Regulation}. However, generalizing their results has remained difficult. 
Our work here establishes a thermodynamically grounded framework that allows a unified description of different crosslinking scenarios.

Earlier theories for filament bundles~\cite{kruse2000actively, kruse2001self, kruse2003self,kruse2005generic}, for suspensions of active filaments~\cite{liverpool2008hydrodynamics, saintillan2008instabilities, de2024supramolecular}, and for densely crosslinked filament networks~\cite{furthauer2019self, furthauer2021design} were formulated using phenomenological constitutive relations for the fluxes or forces acting between filaments and between filaments and motors. Our work extends and complements these approaches by offering a unified framework that derives phenomenological flux and force laws for crosslinked filament pairs. In earlier work, authors considered force velocity rules that balanced active forces against friction with an external medium \cite{liverpool2005bridging, saintillan2008instabilities, de2024supramolecular} or interfilament friction \cite{furthauer2019self,furthauer2021design,linehan2024subcellular}. Our theory provides a basis for delineating the realm of validity of each of these approaches. 
We show that three classes of forces - entropic, active and frictional - emerge between crosslinked filaments and are able to quantify their relative importance from crosslink and filament properties.

In the first part of this paper, we present the main results of our framework. By using ideas from non-equilibrium thermodynamics~\cite{DeGrootPMazur,Kondepudi2015modern,Landau1980statistical,Landau1987fluid} and the hydrodynamics of active gels~\cite{kruse2005generic, furthauer2012taylor, julicher2018hydrodynamic}, we show that the forces on filaments can be written in terms of four contributions: (i) entropic forces, (ii) active forces, (iii) crosslinker-mediated inter-filament friction, and (iv) fluid-mediated friction on the filaments, of which the first three contributions (i-iii) are directly set by crosslinkers. Further, we describe the collective dynamic of the crosslinkers bound to the filaments within the same framework. Most importantly, we give expressions for each of these contributions in terms of measurable properties of crosslinkers and filaments.

In the second part of the paper, we demonstrate the power and utility of this framework by applying it to different experiments involving passive, motorized, and mixtures of motorized and passive crosslinkers, all described within the same unifying language.
We show that for all of the experiments considered here the dynamics of the system can be recovered with our theory using measured parameters from experiments.
For each of the experiments, we discuss the relative importance of inter-filament friction, external friction, active forces and entropic forces. 
We find that for all the experiments considered here, friction against the external medium is negligible compared to internal friction and entropic effects and steady states are set by balancing entropic against active and external forces. This will inform future modeling of filament interaction forces for cytoskeletal networks.

In total, we present a thermodynamically consistent theory for force generation and pattern formation between crosslinked bio-filaments. This theory provides a unifying language and material properties to compare between a variety of crosslinked filament systems. It will inform the quantitative understanding of the cytoskeleton and help the design of engineered polymer networks, by providing prescriptions for the design of crosslink mixtures with desirable mechanical response.

\section{Setup}

\begin{figure}[t]
    \centering
    \includegraphics[width=1\linewidth]{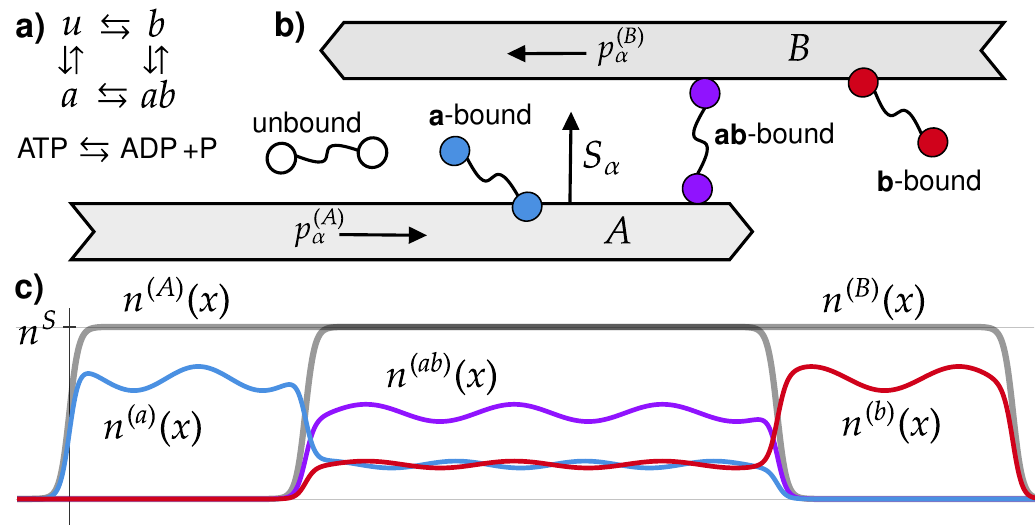}
    \caption{\textbf{Sketch and continuous description of crosslinkers and filament pair.} a) Chemical reaction network: we consider binding reactions between $a, b, ab$, and $u$ states and hydrolysis of ATP to ADP + P. b) Sketch of the system. The filaments $A$ and $B$ are decorated with a-bound, b-bound or ab-bound crosslinkers. Each filament has a polarity vector $p^{(A)}_\alpha$ or $p^{(B)}_\alpha$. The vector perpendicular to the filaments direction pointing from A to B is $S_\alpha$. c) Sketch of the continuous description in one dimension. The filaments ($A, B$) and crosslinkers ($a,b,ab,u$) are described by continuous fields, $n^{(A)}$, $n^{(B)}$, $n^{(a)}$, $n^{(b)}$ and $n^{(ab)}$, respectively. The maximum number density of crosslinkers on filament is $n^S$. }
   \label{fig: description} 
\end{figure}

We study a system consisting of two rigid filaments, denoted $A$ and $B$, which are connected by crosslinkers distributed continuously along their lengths (see Fig.~\ref{fig: description}). In our description, both filaments and crosslinkers are represented by continuous number densities $n^{(i)}$, where the superscript $i$ labels the various crosslink and filament species. Specifically, $n^{(A)}$ and $n^{(B)}$ denote the number densities of the particles that constitute filaments $A$ and $B$, respectively. The space available for crosslinker attachment is taken to be proportional to $n^{(A)}$ and $n^{(B)}$, and for simplicity we set the proportionality constant to unity.
We consider motorized and passive crosslinkers that can be in one of four binding states: unbound, bound to filament $A$, bound to filament $B$, or simultaneously bound to both filaments. 
We denote the corresponding number densities by $n^{(u)}$, $n^{(a)}$, $n^{(b)}$, and $n^{(ab)}$, and in cases of several crosslinker types coexisting we add an additional subscript for the particle type. For instance $n^{(ab)}_{\mathrm{Ase1}}$ will represent the density of doubly bound crosslinkers of the type Ase1. 
Finally, the polarity vectors $p^{(i)}_\alpha$ reflect the orientation of the filament(s) to which a crosslinker is attached. Specifically, $p^{(a)}_\alpha = p^{(A)}_\alpha$ for crosslinkers bound to filament $A$, $p^{(b)}_\alpha = p^{(B)}_\alpha$ for those bound to filament $B$, and $p^{(ab)}_\alpha = (p^{(A)}_\alpha + p^{(B)}_\alpha)/2$ for doubly bound crosslinkers.

\section{Forces on crosslinked filaments.}

The central result of this paper is that the force density on filaments A and B can be written as,
\begin{equation}
     f^{(A,B)}_{\alpha}\!\! =\! \partial_\alpha \Pi^{(A,B)}\! + \mathcal{A}^{(A,B)}_{\alpha}\! 
     +\mathcal{F}^{(A,B)}_{\alpha}\! 
     -\gamma \! \left(V^{(A,B)}_\alpha\!\! -\! v_\alpha\right), \label{eq:force_density_text}
\end{equation}
which we derive in the Methods section. The four terms of Eq.~{(\ref{eq:force_density_text})} have distinct physical meaning: entropic forces, active forces, inter-filament friction, and friction against the external environment.

\textbf{The entropic force} $\partial_\alpha \Pi$ arises since crosslinkers compete for a finite number of binding sites ~\cite{kuvcera2021anillin, Braun2016entropic, Lansky2015adiffusible}. The entropic tension in the filaments is $\Pi^{(A,B)} =- n^{(A,B)}\mu^{(A,B)}_{TS}$ where $\mu^{(i)}_{TS} = -T\partial s/\partial n^{(i)}$ is the entropic contribution to the chemical potential $\mu^{(i)} = \mu_H^{(i)}+ \mu^{(i)}_{TS}$ of species $i$. Here $s$ is the entropy density and $T$ the temperature, and $\mu_H^{(i)}$ is the enthalpic chemical potential which stems from particle interactions. 
The entropy density $s$ depends on the densities of available attachment space, i.e. to $n^{(A,B)}$, and the densities of bound crosslinkers $n^{(a,b,ab)}$. Because $s$ increases when more space becomes accessible, the entropic contribution drives the system toward maximizing the overlap between filaments A and B. Importantly, the value of $s$, and thus the entropic tension $\Pi$ can be derived by counting the number of possible crosslinkers binding configurations; see S.I. section \ref{App:statecounting}. 

\textbf{The active force} $\mathcal{A}_\alpha$ arises when molecular-scale motors convert chemical energy into mechanical work~\cite{Hill1974Formalism,Julicher1997modeling, Geyer2023horizontal}. We show that $\mathcal{A}^{(A,B)}_\alpha = n^{(ab)}_{\mathrm{m}}\:p^{(A,B)}_\alpha \zeta \Delta \mu^{\mathrm{ATP}}$; see  Methods. Importantly $\mathcal A_\alpha$ is proportional to both the number $n^{(ab)}_{\mathrm{m}}$ of motorized crosslinkers within the overlap region and the chemical potential difference $\Delta\mu^{\mathrm{ATP}}$ between the fuel (ATP) and its reaction products (ADP + P). The phenomenological coefficient $\zeta$ characterizes motor force generation. Dimensionally $\zeta\Delta\mu^{\mathrm{ATP}}$ is a force per motor and is linked to the motor stall force. In principle $\zeta\Delta\mu^{\mathrm{ATP}}$ can be measured by setting up an experiment in which motor forces balance against an externally applied force.

\textbf{Inter-filament friction} is generated when crosslinked filaments are sliding relative to each other~\cite{Gaska2020Dashpot}. This inter-filament friction is captured by the term $\mathcal F^{(A,B)} =S_\beta \Sigma_{\alpha \beta}$. Here $S_\beta$ is the vector that points from filament A to filament B (see Fig.~\ref{fig: description}) and $\Sigma_{\alpha\beta}$ is the stress in the bound crosslinkers. On viscous time-scales we show that $\mathcal F^{(A,B)} = \pm\gamma^{\times}n^{(ab)}(V^{(B)}_\alpha - V^{(A)}_\alpha)$ where $V_\alpha^{(A)},V_\alpha^{(B)}$ are the velocities of filaments A and B, respectively, and $\gamma^\times$ is a friction coefficient. 
Here, $n^{(ab)}$ is the concentration of doubly bound crosslinkers. The coefficient $\gamma^{\times}$ can be measured by pulling on filaments. Note that for cases with several crosslinker types each type of crosslinker (motorized or passive) contributes to inter-filament friction. For motorized crosslinkers $\zeta\Delta\mu^{ATP}/\gamma^{\times}$ corresponds to the speed at which otherwise force free filaments would slide past each other.

\textbf{The drag force} between the filament and the surrounding medium is $-\gamma(V_\alpha-v_\alpha)$ where $v_\alpha$ is the local center-of-mass velocity of the system. The drag coefficient $\gamma$ can be calculated from the filament geometry and the properties of the surrounding medium~\cite{Tirado1979translational,Tirado1984comparison,Heyes2019translational}.

Together, these four contributions fully characterize the physical forces acting along pairs of crosslinked filaments. Most importantly, all parameters that enter these forces can be measured experimentally or predicted using structural information on filaments and crosslinkers. Since these forces emerge from crosslinker dynamics, we now turn to examining the collective behavior of crosslinkers along a filament pair.

\section{Crosslinker dynamics}
Each crosslinker density obeys a continuity equation in the form,
\begin{equation}
    \partial_t n^{(i)} = -\partial_{\alpha}\left(j^{(i)}_{\alpha} + n^{(i)} v_\alpha \right)
    +r^{(i)}\label{eq:conservation_number_text},
\end{equation}
where $j^{(i)}_\alpha$ are the spatial fluxes relative to the center of mass and the chemical fluxes $r^{(i)}$ represent source terms due to chemical reactions. 

\textbf{Chemical fluxes $r^{(i)}$}
describe chemical and binding reactions (see Fig.~\ref{fig: description}(a)). These include attachment and detachment from individual filaments, as well as transitions between singly bound and doubly bound states. We will use the capital index $I$ to denote the different reactions allowed in the system.
The source terms obey
\begin{eqnarray}
r^{(i)} = -\sum_I a^{(iI)}\bar{\Lambda}^{(I)}\Delta\mu^{(I)} \simeq\sum_Ik^{(i,I)}(n^{(i,I)}_\star- n^{(i)})\nonumber\\
\label{eq:fluxri_text}
\end{eqnarray}
where $\Delta\mu^{(I)}$ are the chemical potential differences between the species at the nodes linked by chemical reaction $I$, $a^{(iI)}$ are the stoichiometric coefficients, and $\bar\Lambda^{(I)}$ are scalar Onsager coefficients. Note that we ignored crossterms between chemical reactions, and other forces in the system for simplicity, see S.I. section \ref{App:OnsagerCoefficients} for the general expression. By linearizing around the steady state concentrations $n^{(i,I)}_\star$ at which $\Delta\mu^{(I)}=0$, we approximated $r^{(i)}$ in terms of the rate constants $k^{(i,I)}$ which can be measured using standard techniques such as FRAP~\cite{howard2002mechanics, Milo2015number}.

\textbf{Spatial fluxes $j_\alpha^{(i)}=-\Lambda^{(i)}\partial_\alpha{\mu}^{(i)}  + \lambda^{(i)} \Delta\mu^{ATP} p^{(i)}_{\alpha}$} describe the motion of crosslinkers relative to the center of mass of the system.
Importantly,  
$\Lambda^{(i)} = D^{(i)}\chi^{(ii)} / (k_\mathrm{B}T)$, denotes the mobilities. They are linked to the Diffusion coefficients of species $i$ via the fluctuation dissipation theorem, which ensures agreement with Fick's law of diffusion~\cite{Landau1987fluid, Kramer1984Interdiffusion, Kondepudi2015modern, Ortiz2013fluctuating}.
The susceptibilities $\chi^{(ik)}$ are defined as $\chi^{(ik)} = k_\mathrm{\mathrm{B}}T(\partial \mu^{(i)}_{\mathrm{TS}}/\partial n^{(k)})^{-1}$, with $k_\mathrm{B}$ the Boltzmann constant. The susceptibilities $\chi^{(ik)}$ are also called isothermal (cross-)compressibility~\cite{Beijeren83diffusion}.

The second term, containing $\Delta\mu^{\mathrm{ATP}}$, describes directed motion driven by ATP hydrolysis, with $\lambda^{(i)}$ characterizing how the chemical energy input is transformed into motion along the polarity vector $p_\alpha^{(i)}$. See \cite{julicher2018hydrodynamic} and the methods section for a detailed derivation.
In full analogy to the relation between mobility and diffusion, we relate the active transport coefficient $\lambda^{(i)}$ to the force-free velocity $v^{(i)}_0$ of motorized crosslinkers via $v^{(i)}_0 = (\lambda^{(i)}/\chi^{(ii)}) \Delta\mu^{ATP}$. 
 
In dilute systems, the susceptibility $\chi^{(ii)}$  goes to $n^{(i)}$. Thus $\lambda^{(i)}\Delta\mu^{\mathrm{ATP}} \simeq v_0^{(i)}n^{(i)}$ and $\Lambda^{(i)}  k_{\mathrm{B}} T \simeq n^{(i)} D^{(i)}$. In denser regions, $\chi^{(ii)}$ accounts for the crowding effect including traffic-jam like behavior that is particularly present in one-dimensional systems such as a Tonks gas. We give explicit expression in the SI.

Together, these expressions lead us to write the spatial flux of particles as
\begin{align} 
  j^{(i)}_\alpha =& -\sum_{k}D^{(ik)} \partial_\alpha n^{(k)} + \chi^{(ii)}\left(\frac{D^{(ii)}}{k_\mathrm{B} T}\partial_\alpha\mu_H^{(i)} + v_0^{(i)} p^{(i)}_\alpha\right).
\label{eq:pheno_fluxi_text}
\end{align}
Here $D^{(ik)}=D^{(i)}\chi^{(ii)}/\chi^{(ik)}$ is the (cross-)diffusion matrix.
 Most importantly the diffusion coefficients $D^{(i)}$ and the force free velocities $v^{(i)}_0$ can be measured~\cite{Lansky2015adiffusible,Svoboda1994force} and the susceptibilities $\chi^{(ik)}$ can be known from structural information on crosslinkers and the filaments that they bind to; S.I. section \ref{App:statecounting}. The first term of Eq.~({\ref{eq:pheno_fluxi_text}}) encodes the physics of diffusion ($\propto \partial_\alpha n^{(i)}$) and cross-diffusion ($\propto \partial_\alpha n^{(k)}$). The behavior of bound crosslinkers near filament ends is set by the same term via the contributions proportional to $\partial_\alpha n^{(A,B)}$. The second term, proportional to $\partial_\alpha\mu_H^{(i)}$ describes transport in a potential. In the rest of this paper, we will treat cases where $\partial_\alpha\mu_H^{(i)}=0$ for all crosslinkers and motors. This refers to crosslinkers that do not directly interact with one another and whose binding energies to filaments are not dependent on the position along the filament. 
 The last term, proportional to $v_0$ encodes active transport. In SI section~\ref{app:Hill}, we recover Hill's formalism~\cite{Hill1974Formalism} for motors under a load from Eq.~(\ref{eq:pheno_fluxi_text}) in the low density limit.  

Altogether, Eqs.~(\ref{eq:force_density_text}-\ref{eq:pheno_fluxi_text}) describe completely crosslinker dynamics and the forces acting on the filaments in terms of measurable crosslink and filament properties.

\section{Applications}

We next illustrate how our framework provides a unified description of crosslinker-mediated interactions between filaments by applying it to several prototypical experimental situations. Depending on their crosslinker composition, such systems exhibit distinct mechanical behaviors. Filaments connected solely by passive crosslinkers display entropic overlap expansion and inter-filament friction~\cite{kuvcera2021anillin,Gaska2020Dashpot}, whereas motorized crosslinkers alone can generate collective patterns such as traffic jams and exert forces that compete with entropic effects~\cite{Leduc2012,Braun2017changes}. When passive and motorized crosslinkers act together, their interplay can lead to stable sliding dynamics and a pronounced slowing down of microtubules~\cite{Braun2011adaptive}. We begin with the simplest case of an externally driven passive system, where entropic forces dominate the dynamics.

\subsection*{Passive crosslinkers induce entropic contractility in crosslinked bundles}

 \begin{figure*}[t]
     \centering
     \includegraphics[width=1\linewidth]{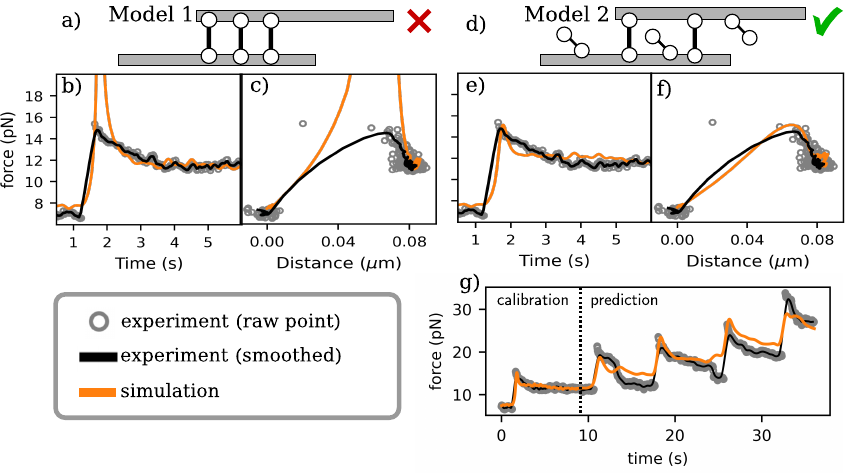}
     \caption{\textbf{Passive crosslinkers induce entropic contractility in crosslinked bundles.} a) Sketch of Model 1 of doubly bound crosslinkers between two filaments. b) Force-Time curves of Model 1 (legend in the gray frame). c) Force-Overlap curves of Model 1. d) Sketch of Model 2 with singly and doubly bound crosslinkers. e) Force-Time curves of Model 2. f) Force-Overlap curves for Model 2. g) Force-Time curves used in e) and for the next four pulling event, done with an independent simulation with the same parameters as in e).}
     \label{fig:Kucera2021}
 \end{figure*}

In the experiment \cite{kuvcera2021anillin} the passive crosslinker anillin has been shown to lead to contractions of filament bundles and rings. By pulling filament bundles apart in steps of 100 nm using an optical trap, the authors showed that the pulling forces increase as the overlap length decreases. We show how these data can be quantitatively explained using our force decomposition Eq.~(\ref{eq:force_density_text}), which allows us to distinguish between two models.

In Model 1 we only consider doubly bound $(ab)$ crosslinkers while in Model 2 we consider $(a,b)$ bound and doubly bound $(ab)$ crosslinkers; see Fig.~{\ref{fig:Kucera2021}(a,d)}. In both cases we will take the system as passive and neglect fluid drag and inter-filament friction, i.e. $v_0 = \gamma = \gamma^\times = 0$. The diffusion coefficient of singly bound crosslinkers has been measured and is $D^{\text{meas}}= 0.0088$ $~\mu$m\textsuperscript{2}s\textsuperscript{-1}, see~\cite{kuvcera2021anillin}.  
We estimate that the diffusion coefficient of doubly bound crosslinkers should be approximately half that value. The maximum number of crosslinkers on filaments is estimated from the length and width of the crosslinkers~\cite{Erickson2009size}. We estimate this number to be approximately 4000 crosslinkers per unit bundle length. This should be interpreted as a coarse estimate, given that the precise size and geometry of the crosslinkers remain uncertain, and that the number of filaments per bundle is itself only approximately known (estimated to be around five~\cite{kuvcera2021anillin}).

The chemical binding and unbinding rates between unbound and singly bound states have been measured in~\cite{kuvcera2021anillin}.
The difference between subsequent plateaus in the force measurement allows to infer the initial overlap length under the assumption that unbinding of singly bound crosslinkers is slow, see S.I. section \ref{app:equations used - sim1}.
The free parameters of the model are the equilibrium concentration of doubly bound crosslinkers, and for model 2 the concentration of singly bound crosslinkers and the transition rate between doubly and singly bound state. The equilibrium concentration is inferred from the initial force measured, see S.I. section \ref{app:equations used - sim1}. The equilibrium concentration of singly bound crosslinkers is set to be small compared to doubly bound concentration, we choose $n^{(ab)}/n^{(a)} =20$ which correspond to a binding energy difference of $4 \: k_\mathrm{B}T$. We take the transition rate between singly and doubly bound to be fast compared to the transition from unbound to singly bound. We choose $k^{(a,b \leftrightharpoons ab)} = 1000\: k^{(u\leftrightharpoons a,b)}$ which allow doubly bound crosslinkers to transiently unbind during pulling event. 

The full equations of motion and all parameter values used are given in SI, and have been solved in a custom code~\cite{GitHubRepo}, built around the Dedalus framework~\cite{Dedalus}.

In our numerical setup we impose the same relative movement of filaments as in experiment and measure the force required to do so using Eq.~(\ref{eq:force_density_text}). Our results are shown in Fig.~{\ref{fig:Kucera2021}} and Supplementary Movie~1.

We first sought to reproduce the force/time and force/distance curves of a single pulling event.
We find that Model 1 is inconsistent with the experimental results. Specifically, the forces massively overshoot in the pulling process; see Fig.~{\ref{fig:Kucera2021}(b-c)}.
In contrast Model 2 describes the data of the pull; see Fig.~{\ref{fig:Kucera2021}(e-f)}.
We conclude, that the dynamic of entropic forces in the system are set by both singly and doubly bound crosslinkers. 

To test the predictive power of this model, we asked whether the parameters calibrated from a single pulling event could accurately predict subsequent pulls. To this end, we performed an independent simulation using the final state and parameters obtained from the first pulling event, and imposed the filament movements as measured in the next four experimental pulls. Remarkably, the simulation reproduces the subsequent pulls; see Fig.~{\ref{fig:Kucera2021}(g)}. This demonstrates that our theory correctly describes entropic forces.

\subsection*{Passive crosslinkers generate inter-filament friction}
Crosslinkers can also generate friction between sliding filaments. To demonstrate how our theory relates to the experimental findings, we will compare to experiment on friction generated by crosslinkers~\cite{Gaska2020Dashpot}. In this experiment, two microtubules coupled by the passive crosslinker PRC1 were pulled against each other and the resulting resisting force was measured. Forces were measured at different velocities and crosslinker numbers. The key finding of~\cite{Gaska2020Dashpot} was a resisting force  that was independent of overlap length.  

\begin{figure*}[t]
    \centering
    \includegraphics[width=1\linewidth]{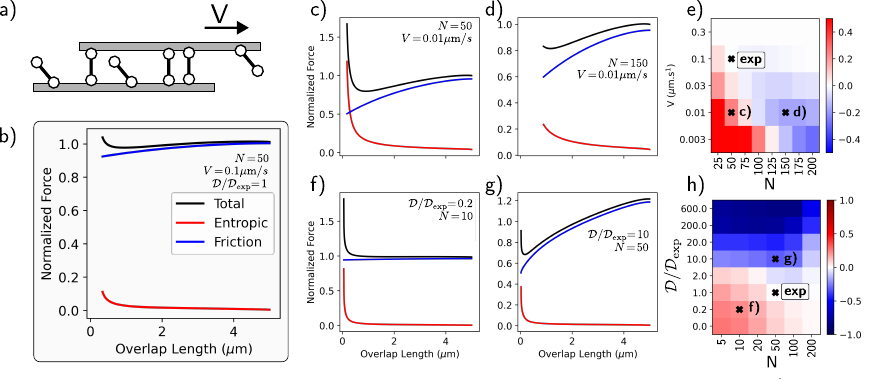}
    \caption{\textbf{Passive crosslinkers generate inter-filament friction.} a) Sketch of the model used in the simulation. b) Normalized Force-Overlap curve obtained in simulation with experimental parameters and V=0.1 $\mu$m-s\textsuperscript{-1}, for N=50 crosslinkers in the overlap. The total force is in black, the entropic and friction contribution are in red and blue, respectively. c) Normalized Force-Overlap curve with V=0.01 $\mu$m-s\textsuperscript{-1} and N=50. d) Normalized Force-Overlap curve with V=0.01 $\mu$m-s\textsuperscript{-1} and N=150. e) Heat map of the normalized total force difference varying the V and N. Crosses indicate the locations of the simulations shown. f-h) At fixed velocity (V=0.1 $\mu$m.s\textsuperscript{-1}), we change the parameter $\mathcal{D}/\mathcal{D}_{\text{exp}}$ and the density. f) Force-overlap curve $\mathcal{D}/\mathcal{D}_{\text{exp}}=0.2$. g) Force-overlap curve $\mathcal{D}/\mathcal{D}_{\text{exp}}=10$. g) Heat map of the normalized total force difference varying the Damköhler number $\mathcal{D}$ and the density.
    }
    \label{fig:Gaska2020}
 \end{figure*}

We model this system by again allowing for singly and doubly bound crosslinkers, see Fig.~\ref{fig:Gaska2020}(a). The inter-filament friction coefficient was measured in experiment and is $\gamma^\times = 2$ pN/($\mu\mathrm{m}.\mathrm{s}^{-1}$). Using parameters from experiments (see S.I. section \ref{tab:parameter used}) we recover the main observation of~\cite{Gaska2020Dashpot} which is that the inter-filaments friction force generated by the crosslinkers is approximately independent of the filaments overlap. Over a numerical pulling experiment the total force (black line) barely varies; see Fig.~{\ref{fig:Gaska2020}(b)} and Supplementary Movie~2.  The force consists of a frictional contribution proportional to $\gamma^\times$, and an entropic one. Only near disconnection the entropic contribution to the force (red) increase, for the reasons discussed in the previous section. Moreover, the friction (blue) slightly decreases from the loss of crosslinkers during pulling. Both of these effects have opposite signs and the total change is negligible in the regime in which the experiments of~\cite{Gaska2020Dashpot} were conducted.   

We next sought to better understand in which regime the observation of constant friction would hold. We perform the same simulation varying the pulling velocity V and the initial number $N$ of crosslinkers in the overlap; see~Fig.~{\ref{fig:Gaska2020}(c-e)}. We measure $\Delta F = F(t=0) -F(t=t_d)$ where $t_0$ is the initial time and $t_d$ is the time at which the filaments disconnect, and $F$ is the force required to pull at the set velocity $V$. Thus, large $\Delta F$ corresponds to a large change in resistive force when pulling the filaments. In the figure Fig.~{\ref{fig:Gaska2020}(e)}, we show the result of this parameter exploration, and find regimes where entropic and/or friction forces become important.  We find that the experimental regime of \cite{Gaska2020Dashpot} requires a sufficiently large pulling velocity. At low crosslinker number and lower velocity, entropic effects start to dominate, since frictional forces are small both from low velocity and low crosslinker number. Conversely, at high crosslinker numbers the loss of crosslinkers via unbinding becomes increasingly important if the pulling is too slow. To avoid this, the time scale of pulling needs to be faster than the characteristic time scale of crosslinker unbinding.

We next investigated the same experiment with different crosslinker binding/unbinding dynamics. We change the Damk\"ohler number $\mathcal{D} = k^{(a,b \leftrightharpoons ab)}L^2/D^{(ab)}$ (where $L = 20$ nm is the typical length of the crosslinker) and the initial crosslinker number $N$. The Damköhler number characterizes the balance between reaction and diffusion time scales. Low values correspond to crosslinkers that unbind slowly, and thus typically diffuse over several crosslinker sizes throughout their life, while large numbers describe crosslinkers which unbind before diffusing over any important length.  In Fig.~{\ref{fig:Gaska2020}(f,g)} we show Force-Overlap curves for different values of $\mathcal{D}$ and number of crosslinkers. The results of this numerical investigation are summarized in Fig.~{\ref{fig:Gaska2020}(h)}. From this we expect deviations from the constant resistive force at large $\mathcal{D}$ or very low densities. Again, the experimental regime of \cite{Gaska2020Dashpot} corresponds to parameters where friction is important enough to dominate over entropic forces and crosslinker unbinding is rare.

These numerical experiments, demonstrate that our theory captures frictional forces correctly, and allows us to place experimental results in a larger space of parameter values. Note also that in this experimental setup crosslink-mediated friction is the dominant contribution, being approximately 200 times larger than fluid-mediated friction. Thus, these experiments are in the highly crosslinked regime \cite{furthauer2022cross}.


 \begin{figure*}[t]
    \centering
    \includegraphics[width=1\linewidth]{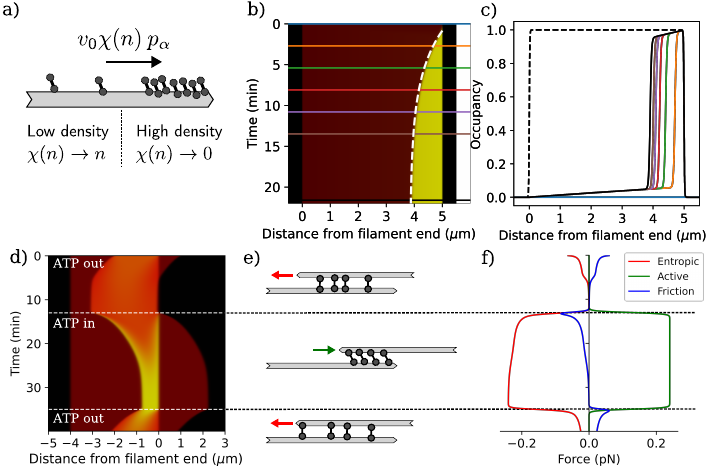}
    \caption{\textbf{Motorized crosslinkers patterning on a single filament and the active force generated between two filaments}  a) Sketch of a single motor walking along a filament under an external load. This represents the low-density limit of our framework, consistent with Hill’s formalism. b) Sketch of a group of motors walking along a single filament. c) Kymograph of the density profile. Color lines correspond to the times at which density profiles are plotted. White dotted curve is the model for the size of traffic jam $(L-a)(1-\exp(-b t))$ with a=3.85 $\mu$m and b=0.003 s\textsuperscript{-1} as measured in \cite{Leduc2012}. d) Density profiles at different times. Color lines are the density of motor, black dashed line is the density of filament A, which is the available attachment space for motors. e) Kymograph of the simulation with motorized crosslinkers between two filaments. The filaments are in red and the motorized crosslinkers in green. White dotted lines represent the change of external ATP load. f) Sketch of the filaments and crosslinkers states during the three phases of the simulation. g) Entropic (red), friction (blue), and active (green) forces during the simulation.} 
    \label{fig:Leduc2012}
 \end{figure*}

\subsection*{The interplay of motor activity and available space determines patterning of motors on single filaments}

We next discuss the behavior of motorized crosslinkers on a single filament. It has been shown experimentally in \cite{Leduc2012} that traffic jams of motorized crosslinkers can occur on filaments. To reproduce this, we model a single species (a) of motorized crosslinkers walking on a single filament (A) (see Fig.~{\ref{fig:Leduc2012}(b)} and Supplementary Movie~3). Importantly, all model parameters were measured in \cite{Leduc2012}; see S.I. \ref{tab:CouplingCoefficients}. With this, Eq.~(\ref{eq:pheno_fluxi_text}) fully determines the dynamics of the system. At high enough density the competition for space on filament balance motor driven fluxes, which are the origin of the observed traffic jams. In our model, the susceptibility is $\chi^{(a)} =n^{(a)}(n^{(A)}-n^{(a)})/n^{(A)}$, we thus retrieve from thermodynamic considerations the model used in~\cite{Leduc2012,Parmeggiani2004totally}. The density profiles that emerge from solving this system are shown in Fig.~{\ref{fig:Leduc2012}(d)}. As in the experiment, motorized crosslinkers accumulate at the filament end. In Fig.~{\ref{fig:Leduc2012}(b-c)} we show that the numerically determined growth of the traffic jam region (color plot) reproduces the measured behavior (dashed white line) exactly. 
This shows that our framework naturally captures concentration-dependent motor velocity relations and the emergence of motorized crosslinker traffic jams.

\subsection*{Motorized crosslinkers between filaments generate stable overlaps by an interplay of active and entropic forces}

We now discuss the behavior of motorized crosslinkers between two filaments, similar to the experimental setup in~\cite{Braun2017changes}, from which we also take our simulation parameters; see S.I. \ref{tab:CouplingCoefficients}. We consider the motorized crosslinkers between two overlapping antiparallel filaments, see Fig.~{\ref{fig:Leduc2012}(e-g)} and Supplementary Movie~4. Motorized crosslinkers can produce an active force on filaments as discussed with Eq.~(\ref{eq:force_density_text}). We simulate a case where the motors are first ATP-starved ($\Delta\mu^{\mathrm{ATP}}=0$ and thus $v_0=0$), constraining them to behave like passive crosslinkers and producing an entropic force which increases the overlap length; see Fig.~{\ref{fig:Leduc2012}(e)}. With the addition of ATP ($\Delta\mu^{\mathrm{ATP}}\neq0$), the crosslinkers become motorized and act against the entropic force, reducing the overlap length to a stable value. Taking away ATP takes the crosslinkers back to their passive state. This demonstrates that our theory captures the competition between motor activity and entropic expansion, which can lead to reversible control of filament overlap. The forces in this scenario are displayed in Fig.~{\ref{fig:Leduc2012}(g)} and highlight the importance of entropic forces for stabilizing non-equilibrium steady states in cytoskeletal structures. To demonstrate this further we next turn to a system in which active and passive crosslinkers coexist.

\subsection*{Motorized and passive crosslinkers compete to define the overall force balance and sliding movement}

 \begin{figure*}[t]
    \centering
    \includegraphics[width=1\linewidth]{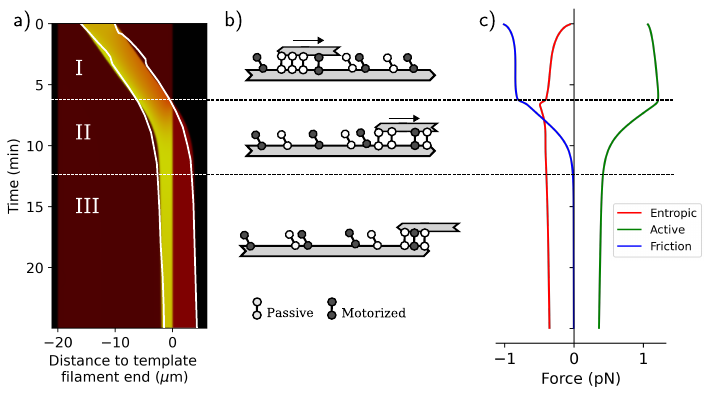}
    \caption{\textbf{Motorized and passive crosslinkers compete to define the overall force balance and sliding movement.} a) Kymograph of the simulation. Template filament is in dark red, sliding filament in red and passive crosslinkers in green. Motorized crosslinkers are not shown. White plain curve is the outline of the experimental observation of \cite{Braun2011adaptive}. White Dashed lines separate the phase I, II, and III. b) Sketches of the system for each phase. c) Plot of the entropic, active and friction forces.}
    \label{fig:Braun2011}
 \end{figure*}

We next investigate a system composed of a mixture of motorized and passive crosslinkers on overlapping filaments, inspired by the experiment~\cite{Braun2011adaptive}. In this experiment, the motor Ncd slides a moving microtubule (MT) past a fixed template microtubule. Furthermore, the passive crosslinking protein Ase1 is also present. The experimentalists identified three distinct phases of the dynamics. In phase (I), the sliding microtubule moves along the template MT at about 30 nm/s. As the sliding MT reaches the end of the template MT the dynamics changes (phase II), and finally comes to a halt (phase III). This behavior is attributed to the changes of motor and passive crosslinker numbers caused by the change in overlap geometry, see Fig.~\ref{fig:Braun2011} and Supplementary Movie 5. We will employ our theory to disentangle the role of entropic, frictional and active forces at play.

To model this system, we use our framework with 6 species: (a,b,ab)-bound passive crosslinkers as well as (a,b,ab)-bound motorized crosslinkers, see Fig.~{\ref{fig:Braun2011}(b)}. We set the velocity of the filament $A$ (template filament) to zero, but the filament $B$ (sliding filament) is free to move and the total force acting on it is null. Most parameters of the model were measured; see S.I. \ref{tab:CouplingCoefficients}.

The only unknown parameters of the model are $\zeta\Delta\mu^{\mathrm{ATP}}$ and $\gamma^{\times}$. For simplicity we will choose the same friction coefficient for Ase1 and Ncd. Further we estimate that $\zeta\Delta\mu^{ATP}/\gamma^\times = 2 v_0$, which corresponds to assuming that at low density Ncd slides filaments at twice its walking speed, which is $v_0 = 100\:\: \mathrm{nm.s}^{-1}$ \cite{Banks2023motor}. Thus effectively $\gamma^\times$ becomes our only fit parameter, which we adjust to reproduce the experimental dynamics measured in \cite{Braun2011adaptive}.

As observed in the experiment (white outline in Fig.~{\ref{fig:Braun2011}(a)}) , the filament $B$ slides (phase I) and slows down to near halting when the crosslinkers accumulate in the overlap (phases II, III). Our framework allows us to explain this behavior in terms of the changing relation between active, entropic, and frictional forces. In phase (I), active forces work against crosslinker friction and entropic forces, generated by the asymmetric accumulation of passive crosslinkers in the trailing edge of the test filament. This predicts that in a hypothetical experiment where all the motors were switched off suddenly in phase I, the test filament would snap back. The entropic force is predicted to create an effective elastic response.   

The change of the entropic force in phase II is due to the rearrangement of crosslinkers in the changing overlap geometry. In this phase the entropic forces increase slightly, and are still dominated by crosslinkers near the trailing edge of the test filament. Simultaneously, the number of motors in the overlap, and thus the motor forces decrease until active and entropic forces reach a near balance. As the filament motion slows to a halt, frictional forces drop and become negligible as phase III is reached; see Fig~{\ref{fig:Braun2011}(c)}. The minimal residual motion in phase III is a consequence of slow unbinding of passive crosslinkers. Note also that throughout the simulation with realistic parameters, the effect of fluid-mediated friction remains small compared to that of crosslinker-mediated friction. Specifically in phase I the ratio of the inter-filament force over the drag force is approximately 75, and near the end of phase three is 35. This implies that the system is highly crosslinked \cite{furthauer2022cross} throughout the experiment.

\section{Discussion}
In this paper, we present a unifying framework to describe the forces generated collectively by passive and motorized crosslinkers between two biopolymer filaments. Using linear non-equilibrium thermodynamics and symmetry considerations, the theory yields a small set of measurable parameters which capture crosslinker binding and transport. From there we arrive at Eq.~(\ref{eq:force_density_text}), which gives the forces that crosslinkers generate between filaments. Importantly, we find that four types of force are important here. First, entropic forces come from crosslinkers competing for available space on filaments; second, active motor forces which capture motors pushing and pulling on filaments; third, inter-filament friction; and fourth, filament friction against the external medium. Thus we provide a way for formulating the inter-filament interaction rules, which underlie theories for cytoskeletal networks \cite{liverpool2005bridging,saintillan2008instabilities,furthauer2019self, furthauer2021design, de2024supramolecular}. Importantly, we give expressions for all forces in terms of measurable filament and crosslinker properties, which in principle allow to fully parametrize our theory from experiment.

Using this unifying language, we show that a single formalism describes crosslinker patterning and the force produced between filaments in many experimental scenarios. We quantitatively describe the forces generated by transient passive crosslinkers between two actin filaments \cite{kuvcera2021anillin} and show them to be entropic in nature; see Fig.~{\ref{fig:Kucera2021}}. We also consider the effect of passive crosslinkers between microtubules \cite{Gaska2020Dashpot}. In this experiment, where filaments are more sparsely decorated, inter-filament friction dominates the dynamics. By studying the dependence of this effect on parameters, we show that the interplay between entropic effects and inter-filament friction can be tuned by modulating the binding properties of crosslinkers; see Fig.~{\ref{fig:Gaska2020}}.   

Turning to motors, 
motor traffic jams \cite{Leduc2012}, motor-induced sliding between two filaments \cite{Braun2017changes}, and mixed motor/passive crosslink behavior \cite{Braun2011adaptive}, can also be explained within our theory. Importantly, our framework allows us to distinguish between entropic, frictional, and active effects. As such, it provides explanatory power and elucidates the physics underlying filament interactions, in terms of simple measurable system properties.

The  physics of transiently crosslinked polymers is also important in other contexts beyond the cytoskeleton. Recently, experimental and numerical work has highlighted the possibility of using transient crosslinkers in polymer recycling \cite{clarke2023dynamic,karmakar2024entropic} and the importance of entropic interaction forces in vitrimers \cite{karmakar2025rouse}. The underlying physics is the same as that described here for the cytoskeletal filaments. Since we can now delineate the link between crosslink mixture properties and the forces that they generate between the polymers that they link, our work is relevant for engineering applications beyond living matter. It can enable the design of crosslink mixtures with desirable properties.

In summary, our framework unifies passive and active crosslinking phenomena into a thermodynamically consistent description. It explains a wide range of observations and provides a practical toolkit to discriminate microscopic mechanisms and guide future experiments on cytoskeletal force generation. Importantly, all parameters of the theory are in principle experimentally accessible. Beyond biophysics, it can guide the engineering of polymer materials, by providing a principled way of thinking about forces between transiently crosslinked polymers.

\section{Methods}
\textbf{Numerical simulation.} The governing equations are integrated numerically in Python using the Dedalus package~\cite{Dedalus}. The scripts used are available on~\cite{GitHubRepo} and the equations solved are shown in the SI appendix.\\

\textbf{Conservation laws.} 
Each particle type (including crosslinkers and filaments) obeys the continuity equation Eq.~(\ref{eq:conservation_number_text}). Note that mass conservation leads to $\sum m^{(i)} j^{(i)}_\alpha = -m^{(0)}j^{(0)}_\alpha $, where $m^{(i)}$ are the masses of a particle $i$ and the index $0$ denote the solvent in which the particles are embedded.

Further our system conserves momentum. Thus, the force balance reads $\partial_\beta \sigma^{\mathrm{tot}}_{\alpha\beta} = 0$, where  $\sigma^{\mathrm{tot}}_{\alpha\beta}=\sigma_{\alpha\beta}^\mathrm{E} + \sigma_{\alpha\beta}$ is the total stress tensor, and $\sigma^\mathrm{E}_{\alpha\beta}$, $ \sigma_{\alpha\beta}$ denote the Ericksen stress and the deviatoric stress, respectively. Einstein's summation convention is implied. This fully defines the conservation laws at play.\\

\textbf{Thermodynamic forces and fluxes.} To get the system dynamics, we follow~\cite{kruse2005generic, furthauer2012taylor, julicher2018hydrodynamic}, and postulate a free energy density $f(g_\alpha, n^{(i)})$ for the system where $g_\alpha$ is the momentum density. With this, we define the chemical potentials $\mu^{(i)} = \partial f/\partial n^{(i)}$.  Translation invariance implies that the hydrostatic Ericksen stress obeys  $\sigma^{\mathrm{E}}_{\alpha\beta}= (f - \mu^{(i)} n^{(i)})\delta_{\alpha\beta}$  and the Gibbs-Duhem relation  $\partial_\beta \sigma^{\mathrm{E}}_{\alpha\beta} = -n^{(i)} \partial_\alpha \mu^{(i)}$; see~\cite{degennes1993physics, furthauer2012taylor, julicher2018hydrodynamic}. The entropy production rate $\dot\Theta$ of the system is, at constant temperature, 
\begin{align}
  T\dot\Theta = \int d^3x\left( \sigma_{\alpha\beta} v_{\alpha\beta}
   - j^{(i)}_\alpha \partial_\alpha \bar{\mu}^{(i)} +
    r^{(I)}\Delta\mu^{(I)} 
    \right), \label{eq:entropy_production}
\end{align}
where $\bar{\mu}^{(i)}\! =\! \mu^{(i)}\!-\!\frac{m^{(i)}}{m^{(0)}} \mu^{(0)}$ is the reduced chemical potentials, $v_{\alpha\beta} = \left( \partial_\alpha v_\beta + \partial_\beta v_\alpha \right)/2$ is the strain rate tensor, and $\Delta{\mu}^{(I)}=\sum_i a^{iI}{\mu}^{(i)}$ is the difference of chemical potential associated with reaction $I$. Note that in what follows, we will take $\mu^{(0)}$ to be constant, which implies $\partial_\alpha \bar{\mu}^{(i)} = \partial_\alpha \mu^{(i)}$, for simplicity.

From the entropy production Eq.~(\ref{eq:entropy_production}) we identify the thermodynamic fluxes $\{\sigma_{\alpha\beta} , j_\alpha^{(i)}, r^{(I)}\}$ and their conjugate thermodynamic forces $\{v_{\alpha\beta}, -\partial_\alpha\mu^{(i)}, \Delta\mu^{(I)}\}$. We assume the near-equilibrium regime where the fluxes follow a linear expansion in the conjugate forces~\cite{Onsager1931reciprocal,DeGrootPMazur,degennes1993physics,Kondepudi2015modern,Landau1980statistical,Landau1987fluid}. Symmetry considerations then determine which couplings between forces and fluxes are allowed. The polarity of the filaments $p^{(A,B)}_\alpha$ and the surface to surface vector $S_\alpha$ break spatial symmetry. The full set of symmetry allowed terms is given in the SI appendix. We only consider a subset of these coefficients to be non-zero, for simplicity.\\

\textbf{Chemical fluxes and reaction rates.} The rates of crosslinker binding/unbinding reactions ($i=a,b,ab,u$) obey $r^{(i)}=-\sum\limits_{I} \overline{\Lambda}^{(I)} a^{(iI)} \Delta \mu^{(I)}$.
We expressed the rates $r^{(i)}$ with the contribution from different reactions $I$, i.e. $r^{(i)} = -\sum_I a^{(iI)}r^{(I)}$, and kept the complexity at its minimum by not including any cross-couplings nor active terms. The phenomenological coefficients $\bar{\Lambda}^{(I)}$ are linked to the binding/unbinding rates. To establish this link, we linearize around the steady state concentration $n^{(i,I)}_\star$ such that the flux through the edge $I$ of the reaction network vanishes~\cite{Kondepudi2015modern}, i.e. $r^{(I)}=0$ when $n^{(i)}=n^{(i,I)}_\star$. With this, the fluxes $r^{(i)}$ are
\begin{eqnarray}
    r^{(i)} &\approx& -\sum_I \overline{\Lambda}^{(I)}a^{(iI)} \frac{\partial \Delta \mu^{(I)}}{\partial n^{(i)}}\bigg|_{n^{(i,I)}_\star} \left(n^{(i)}-n^{(i,I)}_\star\right)\nonumber\\
    &\approx&\sum_I\left( k^{(i,I)}( n^{(i,I)}_\star - n^{(i)}\right), 
\end{eqnarray}
which defines the rates constant $k^{(i,I)}$
\begin{align}
    k^{(i,I)}&= \overline{\Lambda}^{(I)} a^{(iI)} \frac{\partial \Delta \mu^{(I)}}{\partial n^{(i)}}\bigg|_{n^{(i,I)}_\star}.\label{eq:koff}
\end{align}
In this paper, we assume that ATP and ADP densities, and thus the chemical potential difference $\Delta\mu^{\mathrm{ATP}}$, are kept constant by a chemostat, which is typical in cells. In general, if one wishes to describe closed systems, the hydrolysis reaction that converts ATP to ADP +P can be treated analogously to other chemical reactions. Finally, we assume that $r^{(A)}=r^{(B)}=0$, corresponding to chemically inert filaments.\\

\textbf{Diffusive fluxes and active transport.} The fluxes of crosslinkers ($i=a,b,ab$) obey $j^{(i)}_\alpha = -\Lambda^{(i)}\partial_\alpha{\mu}^{(i)}  + \lambda^{(i)} p^{(i)}_{\alpha} \Delta\mu^{\mathrm{ATP}}$.
Here, we didn't allow for coupling between the flux of particles $j^{(i)}_\alpha$ and shear strain $v_{\alpha \beta}$.

The transport coefficients $\Lambda^{(i)}$ and $\lambda^{(i)}$ are linked to the diffusion coefficient $D^{(i)}$ and force free velocity $v_0^{(i)}$ by the relations
\begin{align}
    D^{(i)} = \frac{\Lambda^{(i)}}{\chi^{(ii)}}k_\mathrm{B}T\quad\mathrm{and}\quad
    v_0^{(i)} = \frac{\lambda^{(i)}}{\chi^{(ii)}} \Delta \mu^{\mathrm{ATP}}. 
    \label{eq:fick}
\end{align}

The fluxes of filaments ($i=A,B$) obey
\begin{equation}
  j^{(A,B)}_\alpha = -\Lambda  \partial_\alpha{{\mu}}^{(A,B)}  + \lambda p^{A,B}_\alpha \Delta\mu^{(ATP)} \mp \nu  S_\beta  v_{\alpha \beta}. \label{eq:pheno_fluxAB}
\end{equation}
Here, we include the coupling between filaments fluxes and the strain rate with the coefficient $\nu$. The vector $\ S_\beta$ points from the filament to its binding partner, see Fig.~{\ref{fig: description}(b)}. Note that filament A couples with $S_\beta$, and filament B with $-S_\beta$. We consider that both filaments are identical and set $\Lambda^{(A)}\! = \!\Lambda^{(B)}\!=\! \Lambda$, $\nu^{(A)}=\nu^{(B)}=\nu$, and $\lambda^{(A)}=\lambda^{(B)}=\lambda$. The parameters $\Lambda$, $\lambda$, and $\nu$ scale in the same way as the crosslinker coefficients; this implies that they are proportional to the number of filament particles $n^{(A,B)}$.
Importantly, the coefficient $\lambda$ describes the active flux of the filament generated by energy consuming processes. Then, $\lambda\Delta\mu/\Lambda$ is the force that motors exert locally on the filament. Thus the magnitude of $\lambda$ will also depend on local motor densities. Moreover, the coefficient $\nu$ describes the flux of filaments generated by the strain rate $v_{\alpha\beta}$, so $\nu v_{\alpha\beta}/\Lambda$ is the force that crosslinkers generate from internal shear stress, thus its magnitude will depend on doubly bound crosslinkers density. \\


\textbf{Filament tension.}
The chemical potentials of filaments are
\begin{equation}
    \mu^{(A,B)} = \mu^{(A,B)}_H+\mu^{(A,B)}_{TS}.\label{eq:FilamentsChPot}
\end{equation}
The enthalpic parts $\mu^{(A,B)}_{H}$ contain elastic energies stored within stretched filaments, and are the Lagrange multipliers that enforce the filament in-extensibility constraints. These constraints read
\begin{equation}
    j_\alpha^{(A,B)} +n^{(A,B)}v_\alpha = n^{(A,B)} V_\alpha^{(A,B)}, \label{eq:FilamentsConstraints}
\end{equation}
and ensure that filament particles are convect at filament speeds $V_\alpha^{(A,B)}$. Note that in general $V^{(A,B)}$ can be a function of the position along the filament, for instance when filaments rotate.
Importantly, Eqs.~(\ref{eq:FilamentsChPot}) and the constraints Eqs.~(\ref{eq:FilamentsConstraints}) in Eqs.~(\ref{eq:pheno_fluxAB}), yield two differential equations for $\mu^{(A,B)}_{H}$ which read

\begin{align}
    \partial_{\alpha} \mu_H^{(A,B)} &= - \partial_{\alpha} \mu^{(A,B)}_{TS}
    + \frac{\lambda}{\Lambda}p_\alpha^{(A,B)} \Delta\mu^{\mathrm{ATP}} \nonumber \\
    & \mp \frac{\nu}{\Lambda} S_\beta v_{\alpha \beta} \:-\frac{n^{(A,B)}}{\Lambda} (V^{(A,B)}_\alpha-v_\alpha)
    \label{eq:constraint}
\end{align}
Note that $\partial_\alpha\mu_H^{(A,B)}$ are forces per particle and thus $n^{(A,B)}\mu_H^{(A,B)}$ are the elastic stresses in filaments A and B, respectively.

\textbf{Forces on transiently crosslinked biopolymers.}
In order to get the force density acting on each filament, we multiply the force per particles $\partial_\alpha \mu_H^{(A,B)}$ by the density of filament particles $n^{(A,B)}$. The force density on each filament is then $f^{(A,B)}_\alpha =  n^{(A,B)}\partial_\alpha\mu_H^{(A,B)}$. With this, using equation Eq.~(\ref{eq:constraint}) leads to Eq.~(\ref{eq:force_density_text}) which is the sum of the entropic force $\partial_\alpha \Pi$, the active force $\mathcal{A}_\alpha$, the inter-filament friction force $\mathcal{F}_\alpha$, and the drag force between filament and surrounding fluid. 

The entropic tension is given by
\begin{equation}
\Pi^{(A,B)} =  - n^{(A,B)}\mu^{(A,B)}_{TS}.\label{eq:force pressure} 
\end{equation}

The active force, which is generated by motorized crosslinkers, is
\begin{equation}
    \mathcal{A}^{(A,B)}_{\alpha} = \zeta \Delta\mu^{\mathrm{ATP}}  n^{(ab)}_{\mathrm{m}} p_\alpha .
    \label{eq:force active}
\end{equation}
where we set $(\lambda/\Lambda) n^{(A,B)} = \zeta n^{(ab)}_\mathrm{m}$. Here, we identified the active part of Eq.~(\ref{eq:constraint}) with the standard expression for the force produced by active crosslinkers. This identification allows us to relate the filaments Onsager coefficients $\lambda$ and $\Lambda$, to the experimentally measurable force-generation parameter $\zeta \Delta \mu^{\mathrm{ATP}}$ of motorized crosslinkers.

The shear stress in the crosslinkers that connect the two filaments generates a friction force between the filaments, which is given by
\begin{equation}
    \mathcal{F}^{(A,B)}_{\alpha} = - \gamma^\times n^{(ab)}\left( V^{(B)}_\alpha - V^{(A)}_\alpha \right) ,
    \label{eq:force shear}
\end{equation}
where we used the approximation $v_{\alpha \beta} \approx (V^{(B)}_\alpha - V^{(A)}_\alpha)/d$, with d the inter-filaments distance,  and set $\gamma^{\times}=\nu n^{(A,B)}/(\Lambda d)$. This approximation assumes that the separation between filaments remains constant and small compared to their lengths, so that the velocity field can be expressed as a difference of the filament velocities.
We identified the shear force in Eq.~(\ref{eq:constraint}) with the usual shear force expression. This leads to link the filaments Onsager coefficient $\nu$ and $\Lambda$ to the effective inter-filament friction coefficient $\gamma^\times$ generated by crosslinkers.

Finally, the last contribution of the force is the drag force. We identify the part proportional to the velocity in Eq.~(\ref{eq:constraint}) with the low Reynolds drag force, which sets $\left(n^{(A,B)}\right)^2/\Lambda = \gamma$.

In the overdamped limit, the total force on either filament obeys
\begin{equation}
F^{^+(A,B)}_\alpha + F^{^-(A,B)}_\alpha + c\!\!\!\int \limits_{L^{(A,B)}} \!\!\!\!\!\!\    f^{(A,B)}_\alpha  dl=0,
\label{eq:Force on filament} 
\end{equation}
where $F^{^+(A,B)}_\alpha$ and  $F^{^-(A,B)}_\alpha$ are the forces that external processes apply to the filament tips on their $\pm$ ends, respectively, and $L^{(A,B)}$ are the lengths of the filaments A, B, and $c$ is the cross section of the whole system, which we consider constant. Because we set c constant along the system and assume it is small compared to the filament lengths, the description is effectively one-dimensional. However, the theoretical framework itself is not restricted to this limit and can be readily extended to three-dimensional geometries. In this paper, we discussed situations where filaments are free of external forces and slide just under the effect of motorized crosslinkers ($F^\pm_\alpha = 0$) and situations where filaments are actuated on either end via experimental devices, in which case a force acts on the corresponding filament end.\\

{\bf Acknowledgments:}
This project has been funded by the Vienna Science and Technology Fund (WWTF) [10.47379/VRG20002].  Calculations were performed using supercomputer resources provided by the Austrian Scientific Computing Infrastructure (ASC). This research was supported in part by grant NSF PHY-2309135 to the Kavli Institute for Theoretical Physics (KITP). We thank Daniel Needleman, Peter Foster, and Alexandra Zampetaki for insightful comments on the manuscript.



\bibliographystyle{unsrt} 
\bibliography{ff}


\newpage
\clearpage
\onecolumngrid
\setcounter{secnumdepth}{2}
\hspace*{\fill}
\textbf{Supplementary Information}
\hspace*{\fill}

\phantomsection
\section{Boltzmann Entropy of binding configurations} \label{App:statecounting}
We calculate a Boltzmann entropy for filament binding. The spirit of the calculation that follows is essentially the same as in the classic Flory-Huggins \cite{Flory1942} calculation for mixing entropy in bulk.
We consider two overlapping filaments $A$ and $B$ with $N^{(A)}$ and $N^{(B)}$ total binding sites, respectively. The filament overlap region constitute available site for doubly bound particles. We denote this sites with $N^{(AB)} = \min(N^{(A)}, N^{(B)})$. To determine the Boltzmann entropy $S=k_{\mathrm{B}}\ln\Omega$ of the system, we count the number of possible binding configurations $\Omega$ in which in singly bound ($N^{(a)}$ and $N^{(b)}$) and doubly bound ($N^{(ab)}$) fulfill their respective sites $N^{(A)}$, $N^{(B)}$, and $N^{(AB)}$. Note that when a doubly bound crosslink occupy its site, it takes away possible binding site for singly bound particles. 

In details, we count the number of configuration for $N^{(ab)}$ particles in $N^{(AB)}$ available sites, then $N^{(a)}$ particles in $N^{(A)}-N^{(ab)}$ available site on filament $A$, and also $N^{(b)}$ particles in $N^{(B)}-N^{(ab)}$ available site on filament $B$. With this, the number of possible configurations $\Omega$ is
\begin{align}
    \Omega = \frac{N^{(AB)}!}{N^{(ab)}! (N^{(AB)}-N^{(ab)})!}
    \frac{(N^{(A)}-N^{(ab)})!}{N^{(a)}! (N^{(A)}-N^{(ab)}-N^{(a)})!}
    \frac{(N^{(B)}-N^{(ab)})!}{N^{(b)}! (N^{(B)}-N^{(ab)}-N^{(b)})!}. \label{app:eq:number of states}
\end{align}
Moreover, the energy of the system $U$ contains binding energies from crosslinkers and elastic energies stored within stretched filaments. We suppose that both kinds of energy are proportional to the number of particles and set $U = \sum_i \mu^{(i)}_{H} n^{(i)}$ where $\mu^{(i)}_{H}=\varepsilon^{(i)}$ are the binding energy of crosslinkers ($i=a,b,ab$) or the Lagrangian multiplier of filaments inextensibility ($i=A,B$).

Considering a local volume element $\mathcal{V}$, we assess the entropy density $s=S/\mathcal V$
\begin{align}
     s =& k_{\text{B}} \bigg( n^{(AB)} \ln(n^{(AB)}) - n^{(ab)}\ln(n^{(ab)})
     -(n^{(AB)}-n^{(ab)})\ln(n^{(AB)}-n^{(ab)}) \nonumber \\ \nonumber
     +&(n^{(A)}-n^{(ab)})\ln(n^{(A)}-n^{(ab)}) - n^{(a)}\ln(n^{(a)}) 
      -(n^{(A)}-n^{(ab)}-n^{(a)})\ln(n^{(A)}-n^{(ab)}-n^{(a)}) \nonumber \\
     +&(n^{(B)}-n^{(ab)})\ln(n^{(B)}-n^{(ab)}) - n^{(b)}\ln(n^{(b)})
      - (n^{(B)}-n^{(ab)}-n^{(b)})\ln(n^{(B)}-n^{(ab)}-n^{(b)}) \bigg) \:.\label{app:eq:entropy}
\end{align}
The free energy density is directly $f = \sum_i\mu^{(i)}_H -Ts$. 
We can now assess the chemical potentials $\mu^{(i)}$ and the susceptibilities $\chi^{(ik)}$. Note that the number density of available site for doubly bound $n^{(AB)}$ depends on $n^{(A)}$ and $n^{(B)}$. We introduce the dimensionless filaments functions $\varphi^{(A,B)}$ defined with $n^{(A,B)} = n^S \varphi^{(A,B)}$ with $0\leq \varphi^{(A,B)} \leq 1$ and $n^S$ the typical maximal number of crosslinker on filament. Note that $n^S$ is a constant while $\varphi^{(A,B)}$ are continuous functions. With this, we assume that the density of site for the doubly bound crosslinkers is $n^{(AB)} = n^S \varphi^{(A)} \varphi^{(B)}$.
The chemical potentials are defined as $\mu^{(i)} = \partial f/\partial n^{(i)}$. The chemical potentials of the crosslinkers are
\begin{align}
    \mu^{(a)} &=   \varepsilon^{(a)} +k_BT \ln\left( \frac{n^{(a)}}{n^{(A)} -n^{(ab)} -n^{(a)}} \right), \\ 
    \mu^{(b)} &=  \varepsilon^{(b)} +k_BT \ln\left( \frac{n^{(b)}}{n^{(B)} -n^{(ab)} -n^{(b)}} \right),  \\
    \mu^{(ab)} &=   \varepsilon^{(ab)} +k_BT \left( \ln\left( \frac{n^{(ab)}}{n^{(AB)} -n^{(ab)}} \right) + \ln\left(\frac{n^{(A)} -n^{(ab)}}{n^{(A)} -n^{(ab)}-n^{(a)}}\right) + \ln\left(\frac{n^{(B)} -n^{(ab)}}{n^{(B)} -n^{(ab)}-n^{(b)}}\right) \right),
\end{align}
And the chemical potentials of the filaments are
\begin{align}
    \mu^{(A)} &= \mu^{(A)}_H - k_{\mathrm{B}}T \left( \varphi^{(B)}\ln \left(\frac{n^{(AB)}}{n^{(AB)}-n^{(ab)}} \right) +\ln \left( \frac{n^{(A)}-n^{(ab)}}{n^{(A)}-n^{(ab)}-n^{(a)}} \right) \right),\\
    \mu^{(B)} &= \mu^{(B)}_H - k_{\mathrm{B}}T \left( \varphi^{(A)}\ln \left(\frac{n^{(AB)}}{n^{(AB)}-n^{(ab)}} \right) +\ln \left( \frac{n^{(B)}-n^{(ab)}}{n^{(B)}-n^{(ab)}-n^{(b)}} \right) \right).
\end{align}
The susceptibility are defined as $\chi^{(ik)} = k_\mathrm{B}T (\partial \mu_{TS}^{(i)}/\partial n^{(k)})^{-1}$. We write the most important for what follows : the susceptibilities $\chi^{(ii)}$ of the singly bound crosslinkers ($i=a,b$) and the cross-susceptibilities $\chi^{(ik)}$ between the doubly bound crosslinkers ($i=ab$) and the filaments ($k=A,B$): 
\begin{align}
    \chi^{(aa)} & = \frac{n^{(a)}\left( n^{(A)}-n^{(ab)}-n^{(a)}\right)}{n^{(A)}-n^{(ab)}},\\
    \chi^{(bb)} & = \frac{n^{(b)}\left( n^{(B)}-n^{(ab)}-n^{(b)}\right)}{n^{(B)}-n^{(ab)}},\\
    \chi^{(abA)} & = -\left( \frac{\varphi^{(B)}}{n^{(AB)}-n^{(ab)}} +\frac{n^{(a)}}{(n^{(A)}-n^{(ab)})(n^{(A)}-n^{(ab)}-n^{(a)})}\right)^{-1}, \\
    \chi^{(abB)} & = -\left( \frac{\varphi^{(A)}}{n^{(AB)}-n^{(ab)}} +\frac{n^{(b)}}{(n^{(B)}-n^{(ab)})(n^{(B)}-n^{(ab)}-n^{(b)})}\right)^{-1} .
\end{align}
We highlight here that the information of the free energy $f$ fully determines the chemical potentials $\mu^{(i)}$ and the susceptibilities $\chi^{(ik)}$, and one could in principle derive this quantities even for more complex systems.

\phantomsection
\section{Onsager coefficients and constitutive equations}
\label{App:OnsagerCoefficients}
We expand the thermodynamic fluxes $(-\partial_\alpha \overline{\mu}^{(i)}, \Delta\mu^{(I)}, v_{\alpha \beta})$ in terms of thermodynamic forces $(j^{(i)}, r^{(I)}, \sigma_{\alpha \beta})$ to obtain constitutive equations of the system. In the table~\ref{tab:CouplingCoefficients}, we list all the possible coupling coefficients given the broken symmetries of the system $(p^{(i)}_\alpha ,S_\alpha)$.
\begin{table}[h!]
    \centering
    \renewcommand{\arraystretch}{1.5}
    \begin{tabular}{|c|c|c|c|}\hline
         & $\quad-\partial_\alpha\overline{\mu}^{(j)} \: \: \quad$ & \quad $\Delta\mu^{(J)} \:  \quad $ & $\quad v_{\alpha \beta}\:  \quad$\\\hline
        $\quad j^{(i)}_{\alpha} \:  \quad$ 
        &  $\Lambda^{(ij)}$
        &  $p^{(i)}_{\alpha}\lambda^{(iJ)}$
        &  -$S^{(i)}_{\beta}\nu^{(i)}$\\\hline
        $\quad r^{(I)} \:  \quad$
        &  $p^{(j)}_{\alpha}\lambda^{(Ij)}$
        &  $\overline{\Lambda}^{(IJ)}$
        &  $-\xi^{(I)}_{\alpha\beta}$\\\hline
        $\quad \sigma_{\alpha\beta}\:  \quad$
        &  $S^{(j)}_{\beta}\nu^{(j)}$
        &  $\xi^{(J)}_{\alpha\beta}$
        &  $\eta$\\\hline
    \end{tabular}
    \caption{Table of the Onsager coefficients linking the thermodynamic fluxes to the generalized forces.} 
    \label{tab:CouplingCoefficients}
\end{table}
With these coefficients, the fluxes are, in their most general forms,
\begin{align}
    j^{(i)}_\alpha &= -\Lambda^{(ij)}\partial_\alpha \overline{\mu}^{(j)} + p^{(i)}_\alpha \lambda^{(iJ)}-S_\beta \nu^{(i)}v_{\alpha \beta}, \\
    r^{(I)} &= -p^{(j)}_\alpha  \lambda^{(Ij)} + \overline{\Lambda}^{(IJ)} \Delta\mu^{(J)} - \xi^{(I)}_{\alpha \beta} v_{\alpha \beta}, \\
    \sigma_{\alpha \beta} &= -S_\beta \nu^{(j)}\partial_\alpha \overline{\mu}^{(j)} + \xi^{(J)}_{\alpha \beta} \Delta \mu^{(J)} + \eta v_{\alpha \beta},
\end{align}
where Einstein's summation convention is implied. The matrix $\xi^{(I)}_{\alpha\beta}$ can in principle be constructed from the different broken symmetries $p^{(i)}_\alpha$ and $S_\alpha$.

\section{ Recovering Hill Theory}
\label{app:Hill}

In his seminal 1974 work \cite{Hill1974Formalism} Hill proposed a phenomenological theory for molecular motors. Classic Hill theory can be recovered from our formalism in the limit of low motor density. In Hill's work, a single motor operates against an external force $f_\alpha^{\mathrm{ext}}$; see Fig.~\ref{fig:Leduc2012}(a). In our framework such an external force can be introduced by considering a motor that operates in a potential, such that we can rewrite $\partial_\alpha \bar\mu = \partial_\alpha \bar\mu_{TS} - f_\alpha^{\mathrm{ext}}$. In our formalism, this leads to 
\begin{eqnarray}
j_\alpha = -D \partial_\alpha n + \frac{D\chi}{k_\mathrm{B}T} f^{\mathrm{ext}}_\alpha + v_0\chi p_\alpha.
\end{eqnarray}
If we now consider the case of low motor density, such that $\partial_\alpha n\to 0$ and $\chi \to n$, we can cast this into the form given in Hill's paper \cite{Hill1974Formalism}, and find
\begin{equation}
v_\alpha^{\mathrm{motor}} = \lambda_{11} f_\alpha^{\mathrm{ext}} + \lambda_{12} \Delta\mu,
\end{equation}
where $v_\alpha^{\mathrm{motor}}=j_\alpha/n$, $\lambda_{11} = D/(k_{\mathrm{B}}T)=\Lambda/n$, $\lambda_{12}= \lambda/n$. Hill's second equation gives the rate of fuel consumption and is easily recovered from our formalism and reads
\begin{equation}
    r^{(\mathrm{ATP}\leftrightharpoons \mathrm{ADP+P})}/n = \lambda_{22} \Delta\mu^{\mathrm{ATP}} + \lambda_{12}p_\alpha f_\alpha^{\mathrm{ext}},
\end{equation}
Note that Hill's theory naturally emerges when considering a single state motor molecule. More complex motor models can be formulated by introducing more complex networks of chemical reactions, and multi-state motor models such as the two state models discussed in \cite{Julicher1997modeling} can be formulated by making the appropriate choices for the potentials in which motors exist. Multi-state motor models can deviate substantially from simple Hill behavior, as has been demonstrated in \cite{Lau2007nonequilibrium}.

\section{Numerical methods}
\label{app:numerics}

In this section we discuss the equations of motion solved for the application section. We first discuss how simulations are performed, next how we treat the stiff filaments common to all experiments, and then give details on the individual setups. The parameter values for all simulations are summarized in Table~\ref{tab:parameter used}.  

\subsection{Numerical Framework}
\label{app:numercial framework}
All numerical simulations were performed using the Dedalus spectral framework~\cite{Dedalus}, an open-source Python package designed for solving partial differential equations.
It allows us to solve initial-value problems involving all equations used. In all our simulations, we use periodic boundary conditions applied to the computational domain (i.e., the simulation box, not individual filaments).

\subsection{Treatment of stiff filaments}
\label{app:filaments}
Filaments and crosslinkers are described by continuous fields. To obtain a smooth filament field $n^{(A,B)}$, we use a solution to the Cahn-Hilliard equation
\begin{align}
    0 &= -D\partial^2_x\left(-2n^{(A, B)} +4(n^{(A, B)})^3-6(n^{(A, B)})^2 + G\partial^2_x(n^{(A, B)})\right),
\end{align}
where $\int n^{(A,B)} dx = L$ sets the filament length and the smoothing length scale is set by $\sqrt{G}$. Throughout this paper we use $\sqrt G \approx 0.1\: \mu\mathrm{m} $, which is sufficient to smear the interface out over approximately 10 grid points in our numerical simulations. This value is approximate and may be adjusted between simulations.

In all problems, the filaments are treated as stiff filaments moving at velocity $V^{(A,B)}$, their equations of motion are then $\partial_tn^{(A,B)} = -\partial_x( n^{(A,B)}V^{(A,B)})$. For simulation stability, we instead use
\begin{align}
    \partial_t n^{(A, B)} &= -D\partial^2_x\left(-2n^{(A, B)} +4(n^{(A, B)})^3-6(n^{(A, B)})^2 + G\partial^2_x(n^{(A, B)})\right)-\partial_x\left(n^{(A, B)}V^{(A, B) }\right).
\end{align}
The additional Cahn-Hilliard term in these equations drives numerical errors back towards the desired filament shape, and stays small throughout the simulation. 

\section{Simulations: equations of motion and parameter values and additional information}
\label{app:equations used}

\subsection{Entropic forces increase filament overlap}\label{app:equations used - sim1}
{\bf Model 1:} We consider only doubly bound (ab) passive crosslinkers, which can bind within the filament overlap. The number of accessible micro states is set by Eq.~(\ref{app:eq:number of states}) taking $n^{(a)}=n^{(b)}=0$. With this, we can derive the chemical potentials and susceptibilities of the doubly bound crosslinkers and filaments. The system is then fully defined with the Eqs.~(\ref{eq:force_density_text}-\ref{eq:pheno_fluxi_text}). Note that in this simple model, the "cross-diffusion" coefficients between the crosslinkers ($i=ab$) and the filament ($k=A,B$) take the simple form
\begin{align}
    D^{(ik)} = -D^{(ii)} \frac{n^{(i)}}{n^{(k)}}.
\end{align}
Note the negative sign in the previous equation, which make it clear that the filament ends act as a barrier against crosslinkers diffusion.

{\bf Model 2:} We now have (a),(b),and (ab)-bound crosslinkers. This model is then the same as presented in Section \ref{App:statecounting}, and the results are then directly used to solve the equations Eqs.~(\ref{eq:force_density_text}-\ref{eq:pheno_fluxi_text}).

\begin{figure}[h!]
    \centering
    \includegraphics[width=0.5\linewidth]{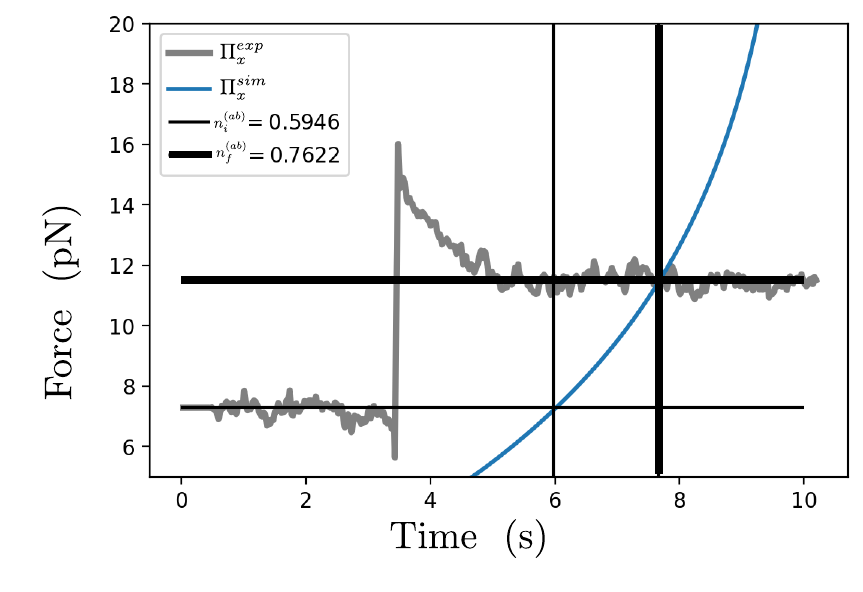}
    \caption{\textbf{Calibration of the initial crosslinker density and initial overlap length.} Grey line is the force measured in experiment. Blue line is the force measured in simulation. Black lines are construction lines to measure when the experimental and simulation forces are equal. Thin black lines correspond to $n^{(ab)} = 0.59\:n^{(AB)}$ and thick black lines correspond to $n^{(ab)} = 0.76\:n^{(AB)}$. This two points correspond to initial and final crosslinker concentrations.}
    \label{app:fig:FitInitialDensity}
\end{figure}

In both model, the initial crosslinker number density $n^{(ab)}_i$ and the initial overlap length $L^o_{i}$ are chosen by measuring the force while varying the doubly bound crosslinker number density. The crosslinker number density is varied manually, and the force measured is compared to the experimental force. The initial and final concentrations of crosslinkers are then measured when the simulation force is equal to the initial or final state experimental force; see Fig.~{(\ref{app:fig:FitInitialDensity})}. The initial overlap length is then calculated by the relation $n^{(ab)}_{i} L^{o}_{i} = n_f^{(ab)}L^{o}_{f}$, using the assumption that there is no loss of crosslinker during one pulling event. We have then $L^o_f =L^o_i+\Delta L$ where $\Delta L$ is the length of the pulling step and is known from the experimental date. This method is applied for the single pulling step and also for the subsequent 4 pulling steps shown in the main text.
In the case of model 2, the concentration of singly bound crosslinkers is chosen to be small compared to the concentration of doubly bound crosslinkers, i.e. ($n^{(ab)}/n^{(a,b)}\approx 20$) which corresponds to a binding energy difference of few $k_{\mathrm{B}}T$ between binding states. We take typically $n^{(ab)} \approx0.5 \:n^{(AB)}$ and $n^{(a)}\approx 0.02\: n^{(A)}$ (same for $n^{(b)}$).



\subsection{Friction between filaments}\label{app:equations used - sim2}
In this simulation, the counting state is the same as in the model 2 in section \ref{app:equations used - sim1}. Moreover, we include the frictional and drag friction forces. The forces on the filaments are then
\begin{align}
    F^{(A)}_x &= c\int \limits_{L^{(A)}} \left(\partial_x\Pi^{(A)}_x - \gamma V^{(A)}_x + \gamma^{\times}n^{(ab)} (V^{(B)}_x-V^{(A)}_x)\right)dl, \\
    F^{(B)}_x &=c\int \limits_{L^{(B)}} \left(\partial_x\Pi^{(B)}_x - \gamma V^{(B)}_x - \gamma^{\times}n^{(ab)} (V^{(B)}_x-V^{(A)}_x)\right) dl.
\end{align}
The total number of crosslinkers $N^{(ab)}$ is measured in experiment. We choose the concentration $n^{(ab)}$ accordingly. For the simulation where we change of the number of doubly bound crosslinkers, we explore a values between $N^{(ab)}=5$ to $N^{(ab)}=200$, which translate into $n^{(ab)} = 0.005\: n^{(AB)}$ and $n^{(ab)} = 0.2\: n^{(AB)}$. Finally, the imposed filament velocities $V^{(A)}$ and $V^{(B)}$ are constant during each simulation until the filament disconnect, i.e. when the simulation crashes because $n^{(ab)}>n^{(AB)}$.

\subsection{The interplay of motor activity and available space determines patterning of motors on single filaments}\label{app:equations used - sim3}

In this model, there is only one type of crosslinkers on a single filament. We then use the results of the section \ref{App:statecounting} with $n^{(B)}=n^{(b)}=n^{(ab)}=0$.
Given the free energy of the system, we derive the chemical potential of the motorized crosslinkers and of the filament,
\begin{align}
    \mu^{(a)} &= \varepsilon^{(a)} + k_\mathrm{B}T \ln \left( \frac{n^{(a)}}{n^{(A)}-n^{(a)}} \right), \\
    \mu^{(A)} &= \mu^{(A)}_H - k_\mathrm{B}T \ln \left( \frac{n^{(A)}}{n^{(A)}-n^{(a)}} \right).
\end{align}
We are particularly interested in the susceptibility $\chi^{(aa)}$ that characterize the active transport of crosslinkers, and the cross term $D^{(aA)}$ which characterize boundary effect between the filament and the crosslinkers,
\begin{align}
    \chi^{(aa)} &= \frac{n^{(a)}\left( n^{(A)}-n^{(a)}\right)}{n^{(A)}}, \\
    D^{(aA)} &= -D^{(a)} \frac{n^{(a)}}{n^{(A)}}.
\end{align}

The spatial flux of the motorized crosslinkers is
\begin{align}
    j^{(a)}_\alpha = -D^{(a)}\partial_\alpha n^{(a)}-D^{(aA)}\partial_\alpha n^{(A)}+v_0^{(a)}\chi^{(aa)}p^{(a)}_\alpha
\end{align}
with $\chi^{(aa)}$ and $D^{(aA)}$ as defined above.

The initial concentration of motorized crosslinkers is set to zero, like in the experiments and simulations~\cite{Leduc2012}.

\subsection{Motorized crosslinkers between filaments can generate stable overlaps by an interplay of motor and entropic forces}\label{app:equations used - sim4}

In this situation, the motorized crosslinkers are now between two filaments, we then use the same equation of the model 1 in section \ref{app:equations used - sim1}, but now include the active and frictional forces. Because filaments are polar but antiparallel, the total polarity is $p^{(A)}_\alpha + p^{(B)} = 0$ and the doubly bound motors do not walk in a particular direction. The total forces on filament A and B are 
\begin{align}
    F^{(A)}_x &= c\int \limits_{L^{(A)}} \bigg(\partial_x\Pi^{(A)} - \gamma V^{(A)}_x +\gamma^{\times}n^{(ab)} (V^{(B)}_x-V^{(A)}_x)  + \zeta \Delta\mu^{(ATP)}n^{(ab)} \bigg) dl,\\
    F^{(B)}_x &= c\int \limits_{L^{(B)}} \bigg( \partial_x\Pi^{(B)} - \gamma V^{(B)}_x - \gamma^{\times}n^{(ab)} (V^{(B)}_x-V^{(A)}_x) - \zeta \Delta\mu^{(ATP)}n^{(ab)}  \bigg) dl.
\end{align}
For this simulations, we choose arbitrarily the number density of motors to be $n^{(ab)} = 0.1\:n^{(AB)}$, this allows us to have entropic, frictional, and actives forces of similar amplitude.

\subsection{Competition between motorized and passive crosslinkers}\label{app:equations used - sim5}
This model considers a more complex version than the section \ref{App:statecounting} with 6 different type of crosslinkers. The number of accessible configuration is 
\begin{align}
    \Omega = &\frac{N^{(AB)}!}{N^{(ab)}_{\mathrm{Ncd}}! N^{(ab)}_{\mathrm{Ase1}}! \left(N^{(AB)}-N^{(ab)}_{\mathrm{Ncd}}-N^{(ab)}_{\mathrm{Ase1}} \right)!} \nonumber\\
    &
    \frac{ \left(N^{(A)}-N^{(ab)}_{\mathrm{Ncd}}-N^{(ab)}_{\mathrm{Ase1}} \right)! }{ N^{(a)}_{\mathrm{Ncd}}!N^{(a)}_{\mathrm{Ase1}}! \left(N^{(A)}-N^{(ab)}_{\mathrm{Ncd}}-N^{(ab)}_{\mathrm{Ase1}} -N^{(a)}_{\mathrm{Ncd}}-N^{(a)}_{\mathrm{Ase1}}\right) }\nonumber\\&
    \frac{ \left(N^{(B)}-N^{(ab)}_{\mathrm{Ncd}}-N^{(ab)}_{\mathrm{Ase1}} \right)! }{ N^{(b)}_{\mathrm{Ncd}}!N^{(b)}_{\mathrm{Ase1}}! \left(N^{(B)}-N^{(ab)}_{\mathrm{Ncd}}-N^{(ab)}_{\mathrm{Ase1}} -N^{(b)}_{\mathrm{Ncd}}-N^{(b)}_{\mathrm{Ase1}}\right) }.
\end{align}
We can then assess the entropy of the system
\begin{align}
     s =& k_{\text{B}} \bigg( n^{(AB)} \ln \left(n^{(AB)} \right) - n^{(ab)}_{\mathrm{Ncd}}\ln \left(n^{(ab)}_{\mathrm{Ncd}}\right)- n^{(ab)}_{\mathrm{Ase1}}\ln \left(n^{(ab)}_{\mathrm{Ase1}}\right)
     -(n^{(AB)}-n^{(ab)}_{\mathrm{Ncd}}-n^{(ab)}_{\mathrm{Ase1}})\ln \left(n^{(AB)}-n^{(ab)}_{\mathrm{Ncd}}-n^{(ab)}_{\mathrm{Ase1}}\right)\nonumber \\ 
     +&(n^{(A)}-n^{(ab)}_{\mathrm{Ncd}}-n^{(ab)}_{\mathrm{Ase1}})\ln \left(n^{(A)}-n^{(ab)}_{\mathrm{Ncd}}-n^{(ab)}_{\mathrm{Ase1}}\right) - n^{(a)}_{\mathrm{Ncd}}\ln \left(n^{(a)}_{\mathrm{Ncd}}\right) - n^{(a)}_{\mathrm{Ase1}}\ln \left(n^{(a)}_{\mathrm{Ase1}}\right) \nonumber \\
      -&(n^{(A)}-n^{(ab)}_{\mathrm{Ncd}}-n^{(ab)}_{\mathrm{Ase1}}-n^{(a)}_{\mathrm{Ncd}}-n^{(a)}_{\mathrm{Ase1}})\ln \left(n^{(A)}-n^{(ab)}_{\mathrm{Ncd}}-n^{(ab)}_{\mathrm{Ase1}}-n^{(a)}_{\mathrm{Ncd}}-n^{(a)}_{\mathrm{Ase1}}\right) \nonumber \\
     +&(n^{(B)}-n^{(ab)}_{\mathrm{Ncd}}-n^{(ab)}_{\mathrm{Ase1}})\ln \left(n^{(B)}-n^{(ab)}_{\mathrm{Ncd}}-n^{(ab)}_{\mathrm{Ase1}}\right) - n^{(b)}_{\mathrm{Ncd}}\ln \left(n^{(b)}_{\mathrm{Ncd}}\right) - n^{(b)}_{\mathrm{Ase1}}\ln \left(n^{(b)}_{\mathrm{Ase1}}\right) \nonumber \\
      -&(n^{(B)}-n^{(ab)}_{\mathrm{Ncd}}-n^{(ab)}_{\mathrm{Ase1}}-n^{(b)}_{\mathrm{Ncd}}-n^{(b)}_{\mathrm{Ase1}})\ln \left(n^{(B)}-n^{(ab)}_{\mathrm{Ncd}}-n^{(ab)}_{\mathrm{Ase1}}-n^{(b)}_{\mathrm{Ncd}}-n^{(b)}_{\mathrm{Ase1}}\right)    
      \bigg) \:,\label{app:eq:entropyMix}
\end{align}
which is used to calculate each chemical potential. This fully defines the equation of motion of the six species.
The filament A is fixed, meanwhile the filament B is free to move. The sliding velocity of the filament B is then calculated from $\int_{L^{B}} f_\alpha^{(B)} dl = 0$. This leads to
\begin{align}
    V_x^{(B)} = \frac{1}{\int \limits_{L^{(B)}}\left(\gamma + \gamma^{\times}\left(n^{(ab)}_{\mathrm{Ase1}}+n^{(ab)}_{\mathrm{Ncd}}\right)\right)dl} \left( \: \int \limits_{L^{(B)}} \!\!\left( n^{(B)} \partial_x \overline{\mu}^{B} + \zeta \Delta\mu^{ATP}n^{(ab)}_{\mathrm{Ncd}}\right) dl \right).
\end{align}
Based on the observation in~\cite{Braun2011adaptive}, the motors Ncd do not accumulate in the overlap region of filaments, while the passive crosslinkers Ase1 accumulate. This means that Ncd do not gain much energy by binding to both filaments, while Ase1 does. We thus set
$\Delta\varepsilon_{\mathrm{Ncd}}\approx1.5\: k_{\mathrm{B}}T$ which leads to a ratio of 2 between the doubly bound and singly bound motors. The difference of energy for Ase1 passive crosslinkers is $\Delta\varepsilon_{\mathrm{Ase1}}\approx3\: k_{\mathrm{B}}T$ which leads to a ratio of 10.


\begin{table}[h]
    \centering
    \caption{Parameters used in the simulations}
    \label{tab:parameter used}
    \begin{tabular}{lp{7cm}p{3cm}p{7cm}}
        \textbf{Description} && \textbf{Value} & \textbf{Notes} \\
        \hline \\
        \multicolumn{4}{l}{\textbf{Entropic forces increase filament overlap}} \\
        $D^{(a,b)}$ & singly bound anillin diffusion coefficient & 0.0088 $\mu\mathrm{m}^2$s\textsuperscript{-1} & \cite{kuvcera2021anillin} \\
        $D^{(ab)}$ & doubly bound anillin diffusion coefficient & 0.0044 $\mu\mathrm{m}^2$s\textsuperscript{-1} & Estimated to be $D^{(a,b)}/2$ \\
        $k^{u\leftrightharpoons a,b}$ & single bound unbinding rate &  0.06 s\textsuperscript{-1} &  \cite{kuvcera2021anillin} \\
        $k^{a,b\leftrightharpoons ab}$ & doubly bound unbinding rate &  60 s\textsuperscript{-1} & 1000 times single bound unbinding rate, from experiment fitting\\
        $\Delta \varepsilon$ & difference of energy between (a,b) and (ab) &  4 $k_\mathrm{B}T$ & leads to a ratio of $\approx20$ between doubly bound and singly bound crosslinkers\\
        $n^S$ & maximal concentration of crosslinkers &  4000$\:\mu$m\textsuperscript{-1} & Estimation from bundle of 5 filaments, for elongated proteins of a size of 124 KDa~\cite{Erickson2009size}. \\
        \hline \\
        \multicolumn{4}{l}{\textbf{Friction between filaments}} \\
        $D^{(a,b)}$ & singly bound PRC1 diffusion coefficient &  0.13 $\mu\mathrm{m}^2$s\textsuperscript{-1}&  \cite{Wildenberg2011single,Gaska2020Dashpot}  \\
        $D^{(ab)}$ & doubly bound PRC1 diffusion coefficient &  0.04 $\mu\mathrm{m}^2$s\textsuperscript{-1} &  \cite{Wildenberg2011single,Gaska2020Dashpot}  \\
        $k^{(u \leftrightharpoons a,b)}$ & single bound unbinding rate &  0.002 s\textsuperscript{-1}  &  value used in simulation in \cite{Gaska2020Dashpot} \\
        $k^{(a,b \leftrightharpoons ab)}$ & doubly bound unbinding rate &  0.002 s\textsuperscript{-1}  &  value used in simulation in \cite{Gaska2020Dashpot} \\
        $\Delta \varepsilon$ & difference of energy between (a,b) and (ab) &  2 $k_\mathrm{B}T$ & leads to a ratio of $\approx5$ between doubly bound and singly bound crosslinkers\\
        $\gamma c L$ & drag coefficient between filament and solvent &  0.4 pN/($\mu$m.s\textsuperscript{-1}) & Estimated for a MT sized cylinder in solvent 100 more viscous than water \\
        $\gamma^{\times}$ & effective drag coefficient per crosslinkers  &  2 pN/($\mu$m.s\textsuperscript{-1}) & $\sim$ 0.2 pN/crosslinkers at V=0.1 $\mu$m.s\textsuperscript{-1} \cite{Gaska2020Dashpot}. The friction created by one crosslinkers is five time the friction between filament and surrounding fluid ($\gamma^{\times}=5\gamma c L$).\\
        
        $n^S$ & maximal concentration of crosslinkers &  200$\:\mu$m\textsuperscript{-1} & minimal spacing of 5 nm between crosslinkers \\
        
        \hline \\
        \multicolumn{4}{l}{\textbf{The interplay of motor forces and entropy on single filament}} \\
        $D$ & Diffusion coefficient &  0.001 $\mu$m\textsuperscript{2}.s\textsuperscript{-1} &  \cite{Bormuth2009friction} \\ 
        $v_0$ & preferred velocity of motors &  0.05 $\mu$m.s\textsuperscript{-1} &  \cite{Leduc2012} \\ 
        $k^{(u\leftrightharpoons a)}$ & unbinding rate &  0.027 s\textsuperscript{-1} & \cite{Leduc2012} \\ 
        $n^S$ & maximal concentration of crosslinkers &  200$\:\mu$m\textsuperscript{-1} & minimal spacing of 5 nm between crosslinkers \\
                
        \hline \\
        \multicolumn{4}{l}{\textbf{Motorized crosslinkers between filaments can generate stable overlaps by an interplay of motor and entropic forces}} \\
        $D^{(ab)}$ & Diffusion coefficient &   0.05 $\mu$m\textsuperscript{2}.s\textsuperscript{-1} & $\simeq$ value from \cite{Braun2017changes}. \\ 
        $\zeta \Delta\mu^{ATP}$ & force per crosslinkers &  8 10\textsuperscript{-3} pN  & \\ 
        $\gamma^{\times}$ & effective drag coefficient from crosslinkers &  0.8 pN/($\mu$m.s\textsuperscript{-1})  &   \\ 
        $\gamma c L$ & drag coefficient between filament and solvent &  0.4 pN/($\mu$m.s\textsuperscript{-1}) & Estimated for a MT sized cylinder in solvent 100 more viscous than water \\ 
        $n^S$ & maximal concentration of crosslinkers &  200$\:\mu$m\textsuperscript{-1} & minimal spacing of 5 nm between crosslinkers \\

        \hline \\
        \multicolumn{4}{l}{\textbf{Competition between motorized and passive crosslinkers}} \\
        $D^{(a,b)}_{\mathrm{Ase1}}$ & singly bound Ase1 diffusion coefficient &  0.09 $\mu\mathrm{m}^2$.s\textsuperscript{-1} &  \cite{Lansky2015adiffusible} \\ 
        $D^{(ab)}_{\mathrm{Ase1}}$ & doubly bound Ase1 diffusion coefficient &  0.011 $\mu\mathrm{m}^2$.s\textsuperscript{-1} &  \cite{Lansky2015adiffusible} \\ 
        $D^{(a,b)}_{\mathrm{Ncd}}$ & singly bound Ncd motor diffusion coefficient &  0.05 $\mu\mathrm{m}^2$.s\textsuperscript{-1} &  \cite{Furuta2008minus}\\
        $D^{(ab)}_{\mathrm{Ncd}}$ & doubly bound Ncd motor diffusion coefficient &  0.025 $\mu\mathrm{m}^2$.s\textsuperscript{-1} &  $D^{(a,b)}_{\mathrm{Ncd}}/2$ \\
        $k^{(u \leftrightharpoons a,b)}$ & Singly bound Ase1 unbinding rate &  0.0017 s\textsuperscript{-1} &  \cite{Braun2011adaptive}\\
        $k^{(a,b \leftrightharpoons ab)}$ & Doubly bound Ase1 unbinding rate &  1.7 s\textsuperscript{-1} & Same ratio as in sim 1.\\
        $\Delta \varepsilon_{\mathrm{Ase1}}$ & difference of energy between (a,b) and (ab) &  3 $k_\mathrm{B}T$ & leads to a ratio of $\approx10$ between doubly bound and singly bound crosslinkers\\
        $k^{(u \leftrightharpoons a,b)}$ & Singly bound Ncd unbinding rate &  0.01 s\textsuperscript{-1} & Chosen that in does not accumulate in the overlap based on the observation of \cite{Braun2011adaptive} \\
        $k^{(a,b \leftrightharpoons ab)}$ & Doubly bound Ncd unbinding rate &  30 s\textsuperscript{-1} & Chosen that in does not accumulate in the overlap based on the observation of \cite{Braun2011adaptive}\\
        $\Delta \varepsilon_{\mathrm{Ncd}}$ & difference of energy between (a,b) and (ab) &  1.5 $k_\mathrm{B}T$ & leads to a ratio of $\approx2$ between doubly bound and singly bound crosslinkers\\
        $v_0$ & preferred velocity of motors &  0.1 $\mu$m.s\textsuperscript{-1} &  \\ 
        $\gamma cL$ & drag coefficient between filament and solvent &  0.4 pN/($\mu$m.s\textsuperscript{-1}) &  Estimated for a MT sized cylinder in solvent 100 more viscous than water  \\
        $\gamma^{\times}$ & effective drag coefficient from crosslinkers & 0.1 pN/($\mu$m.s\textsuperscript{-1}) & Fitting parameter  \\

        $\zeta \Delta\mu^{(ATP)}$ & force per crosslinkers&  2 10\textsuperscript{-2} pN &  \\
        $n^S$ & maximal concentration of crosslinkers &  200$\:\mu$m\textsuperscript{-1} & minimal spacing of 5 nm between crosslinkers \\

    \end{tabular}
\end{table}

\section{Supplementary videos}

For each simulation, we plot the distributions of implied filaments and crosslinkers. The videos of the plots over time can be found in the video section of GitHub repository \cite{GitHubRepo}.\\

\noindent
\textbf{Supplementary video 1} Plots of number densities of the 4 last pulling events shown in Fig.~{\ref{fig:Kucera2021}(g)}. \\
\textbf{Supplementary video 2} Plots of number densities of simulation with the parameters corresponding to experimental values shown in Fig.~{\ref{fig:Gaska2020}(b)}. \\
\textbf{Supplementary video 3} Plots of number densities shown in Fig.~{\ref{fig:Leduc2012}(d)}. \\
\textbf{Supplementary video 4} Plots of number densities of the simulation whose kymograph is shown in Fig.\ref{fig:Leduc2012}(e)\\
\textbf{Supplementary video 5} Plots of number densities of the simulation whose kymograph is shown in Fig.\ref{fig:Braun2011}(a). Passive crosslinkers are denoted by the subscript $p$ and the motorized crosslinkers by the subscript $m$. \\

\end{document}